\documentclass[a4paper,12pt]{article}

\usepackage{amssymb,latexsym,amsmath}

\usepackage{fullpage}
\usepackage{nicefrac}
\usepackage{mathrsfs}
\usepackage{slashed}
\usepackage{bbold}
\usepackage{booktabs}
\usepackage{tikz,pgf}
\usetikzlibrary{arrows,angles,calc,quotes} 

\makeatletter \@addtoreset{equation}{section}
\makeatother

\begin{document}

\begin{titlepage}
	\vspace{35pt}

	\begin{center}
	    { \LARGE{\bf Supersymmetry-breaking compactifications \\[3mm] 
		on Riemann-flat manifolds
}}

		\vspace{50pt}

		{Gianguido Dall'Agata and Fabio Zwirner}

		\vspace{25pt}

		{
		{\it  Dipartimento di Fisica e Astronomia ``Galileo Galilei''\\
		Universit\`a di Padova, Via Marzolo 8, 35131 Padova, Italy}

		\vspace{15pt}

			    {\it   INFN, Sezione di Padova \\
		Via Marzolo 8, 35131 Padova, Italy}
				}

		\vspace{40pt}

		{ABSTRACT}

		\vspace{40pt}


	\end{center}

	\vspace{10pt}

We consider compactifications of higher-dimensional supergravities on Riemann-flat manifolds of dimension $d$ ($d \le 7$) that fully break supersymmetry at the classical level on a resulting $D$-dimensional Minkowski space.
We systematically discuss consistency conditions, the Kaluza--Klein (KK) spectrum and harmonics, and the resulting one-loop effective potential $V_1$, focusing for illustration on maximal supergravity and $d=3$, in particular on the ${\mathbb T}^3 / {\mathbb Z}_3$ and on the Hantzsche-Wendt manifolds.
We show how the KK spectrum can be re-organized in multiplets of the broken $D$-dimensional supersymmetry, derive new universal supertrace mass relations valid at each redefined KK level and obtain an analytic finite expression for $V_1$ after resumming the contributions of all KK levels.
In all examples $V_1$ is negative definite and scales with inverse powers of some internal radii. 
We extensively comment, when applicable, on the relation with the Scherk--Schwarz mechanism and with supersymmetry-breaking string compactifications on freely acting symmetric orbifolds. 
We also finally clarify the assumptions and constraints under which Scherk--Schwarz reductions correspond to twisted tori compactifications.

\end{titlepage}

\baselineskip 6 mm

\renewcommand{\arraystretch}{1.2}

\section{Introduction} 
\label{sec:introduction}

Space-time local supersymmetry seems to be deeply rooted in string theory, our best candidate for a unified quantum theory of the fundamental interactions, but we are still lacking a theoretically and phenomenologically satisfactory realisation in the broken phase.
The expectation that supersymmetry would be realised already in a four-dimensional effective quantum field theory at the Fermi scale of weak interactions has not been borne out by experiment.
It is therefore a good moment to explore alternative possibilities for the scale of supersymmetry breaking.
An interesting one is that supersymmetry is not broken directly at the string scale, close to the Planck scale, but instead at a lower compactification scale, at which a higher-dimensional supergravity is a valid effective field theory approximation, with exponentially suppressed string corrections.

The idea of breaking supersymmetry in flat space by making use of extra compact dimensions dates back to the generalized dimensional reductions of higher-dimensional supergravities introduced by Scherk and Schwarz in \cite{Scherk:1978ta,Scherk:1979zr,Cremmer:1979uq}. 
These models are specific realisations of no-scale supergravity models (for a recent review and references, see e.g. \cite{Zwirner:2025ohv}), in the sense that the scale of the supersymmetry-breaking masses slides along some flat directions of the classical potential, associated with the size of the compact space. 
However, dimensional reduction does not capture the full physical information of the compactified higher-dimensional theory, where the supersymmetry-breaking parameters are quantized: the reduced theory is generically a consistent truncation rather than a  genuine effective theory (see, e.g., \cite{DallAgata:2005zlf,Grana:2013ila}). 
For example, if the compactification manifold is a torus, supersymmetry can appear to be broken in the reduced truncated theory, but it is left unbroken in the full compactified theory.
Moreover, even when supersymmetry is broken, the full compactified theory may contain states with comparable masses, which in part are truncated away and in part included in the reduced theory.     

String models where compactification plays a crucial role in the breaking of supersymmetry (to be contrasted with other non-supersymmetric string constructions where supersymmetry is completely broken at the string scale \cite{Dixon:1986iz,Alvarez-Gaume:1986ghj}) were introduced in \cite{Rohm:1983aq,Ferrara:1987es,Kounnas:1988ye,Ferrara:1988jx}.
These models contain infinite towers of states that include not only  the Kaluza--Klein states, but also winding modes, corresponding to strings wrapped around the compact space. 
Also these models have classical flat directions associated with the volume of the compact space and controlling the scale of supersymmetry breaking, but we still lack a viable stabilization mechanism.  
Moreover, these models often have tachyons, associated with winding modes, but they can be removed from the spectrum by going to sufficiently large volume.
In fact, winding modes should decouple exponentially fast in the large volume limit and hence allow us to have better control on the effective theory.
However, these models typically also involve other sources of supersymmetry breaking (orbifold or orientifold projections), which introduce localized defects such as orbifold fixed points, D-branes, O-planes, giving rise to additional non-perturbative states and invalidating the field-theory description.
Interesting exceptions, like certain freely acting orbifolds, can be used to draw an accurate comparison with the supergravity results, without the need of introducing additional stringy states, and this therefore is the vantage point we want to take in this work.
Actually, a supergravity analysis is useful for many reasons. 
For instance, if we consider eleven-dimensional supergravity, which does not have a perturbative fully calculable equivalent in string theory, supergravity is the main way to get information on the low-energy behaviour of the model.
Field theory computations in this framework can show that the 1-loop potential is finite and calculable, in the appropriate limit, with no need for the string regulator and no explicit appearance of the string scale in the result and hence show the validity of this approach.

Recently,  we performed in \cite{DallAgata:2024ijh} a systematic analysis of the spectrum and of the one-loop effective potential $V_1$ of pure five-dimensional supergravities compactified on the circle, with supersymmetry fully broken by a Scherk--Schwarz twist of the periodicity conditions. 
Taking advantage  of the very simple form of the KK spectrum, we observed that the supertraces of the field-dependent mass matrices are the same for each KK level, and we confirmed that, after resumming over all KK levels, $V_1$ is finite for any number $N > 0$ of supersymmetries. 
In particular, in the case of $N=8$ supergravity, we found results that extended but qualitatively confirmed those obtained for the four dimensional theory \cite{DallAgata:2012tne}, i.e. that at any KK level $n$ on the circle, after accounting for the supersymmetry-breaking mass shifts, ${\rm Str} {\cal M}_{n}^{2p} = 0$  for $p<4$, and ${\rm Str} {\cal M}_{n}^{2p} > 0$ (but finite) for $p=4$.
Also, we confirmed the finding of \cite{DallAgata:2012tne} that $V_1$ is negative definite even after resumming the contributions from the KK states.  

In this work we further extend these results by analyzing KK compactifications on Riemann flat manifolds of dimension $d$ ($d \le 7$). 
The reason for this choice is that compactifications on these manifolds contain (but do not exhaust) the Scherk--Schwarz compactifications that can be obtained by focusing on the fluctuations on the background values determined by the identity structure of twisted tori.
In all cases the truncated theory has the original supersymmetry at the Lagrangian level and supersymmetry is broken spontaneously.
By the procedure described in this work we can obtain the full KK spectrum on these same manifolds and hence we can better understand supersymmetry breaking and how this applies to the full tower of states and not just to the truncated theory.
While there were some attempts at discussing the KK spectrum of a Scherk--Schwarz compactification \cite{Andriot:2018tmb,Gkountoumis:2023fym}, this was taking into account only a subset of all states.
On the other hand, other authors discussed the full string theory spectrum on some of these manifolds \cite{Acharya:2020hsc}, but there was neither a clear connection with Scherk--Schwarz compactifications nor a clear identification of how the underlying supersymmetry affects the properties of the spectrum and constrains the supersymmetry breaking mechanism.

Here we offer a thorough supergravity analysis of such backgrounds and provide a clear algorithm to explicitly write the harmonics on Riemann-flat internal manifolds, and to compute their KK spectrum.
While doing this we find some remarkable identities on the supertraces of the field-dependent mass matrices valid at each KK level, once the states are suitably re-organized in multiplets of the broken $D$-dimensioal supersymmetry: this extends the results of \cite{DallAgata:2012tne} and \cite{DallAgata:2024ijh} to arbitrary d-dimensional Riemann-flat manifolds.
In detail, we find that for theories with maximal supersymmetry ${\rm Str} {\cal M}^{2p} = 0$, at each redefined KK level, for any $p<4$, while for $p=4$ we find a non-zero positive value independent of the KK level.
It must be stressed that this is a rather non-trivial result, which implies (\emph{but is not implied by}) the well-known fact that summing over all KK states one always finds vanishing results (for instance for a single-index tower, $\zeta(-2n) = 0$, $\forall n \in {\mathbb N}$).
It is in fact a signal of the spontaneous breaking of supersymmetry, which rearranges states in multiplets of D-dimensional supersymmetry even after supersymmetry is broken, though with different masses.
This is particolarly remarkable for the example of the Hantzsche--Wendt (HW) manifold, because such a manifold does not admit an ordinary interpretation as a twisted torus (it is the fibration of a Klein bottle over an interval), nor as a consistent Scherk--Schwarz compactification.

Our general algorithm also allows us to understand the general structure of the spectrum from first principles. 
Every time one has a holonomy group that does not allow sub-maximal orbits for certain values of the KK numbers and a trivial spin structure exists, the corresponding spectrum is fully supersymmetric.
When sub-maximal orbits exist, on the other hand, the spectrum is still organized in multiplets, but it is non-degenerate.
In particular, fermions exhibit non-symmetric spectra.

Once we have the full spectrum, we can obtain the 1-loop $D$-dimensional scalar potential $V_1$ and we provide a general formula for it, based on the Epstein zeta function \cite{epsteinzeta}.
This essentially generalizes available field theory computations of the Casimir energy on these manifolds and can be used as a starting point for a systematic study of the 1-loop corrections to the classical Minkowski vacuum for all such manifolds.
While we only look at some simple instances of internal manifolds in this work, because we want to focus on the general structure and theory rather than on a systematic scan of all possibilities, we see once more that such contribution is negative definite for all our examples and inversely proportional to the size of the internal manifold, hence giving a runaway potential.
While in all our examples this may be linked to the observation that one should expect this for backgrounds with more bosonic than fermionic zero modes \cite{Acharya:2020hsc}, we have no general proof yet.

A crucial contribution of our analysis is that we finally obtain an explicit clear understanding of the relation between the Scherk--Schwarz mechanism, twisted tori  and the compactifications on Riemann-flat manifolds.
In particular we show how the selection of some specific states in the full KK spectrum of Riemann-flat manifolds gives those surviving the consistent truncation in the Scherk--Schwarz mechanism.
We give the explicit identification of the modes in the two different frameworks and hence finally clarify the issues raised in \cite{DallAgata:2005zlf,Grana:2013ila}, related to the difference between twisted tori and simple tori when considering truncated or dimensionally reduced theories.

\bigskip

\textbf{Note added:} During the completion of this project, we became aware of \cite{Bento}, which also analyzes the Casimir energy on Riemann-flat manifolds, though for a different purpose.


\section{Spin Bieberbach manifolds} 
\label{sec:spin_bieberbach_manifolds}

Riemann-flat, compact, spin manifolds are a subset of Bieberbach manifolds, a class of manifolds classified by their fundamental groups, which are a subset of crystallographic groups \cite{Charlap1986}.
All $d$-dimensional Bieberbach manifolds $M_d$ can be obtained as freely acting quotients of ${\mathbb R}^d$ with a cocompact (i.e. such that $ {\mathbb R}^d/\Gamma$ is compact) torsion-free~\footnote{A torsion-free group  $\Gamma$ is a group whose only element of finite order is the identitiy $e$, i.e. $g^n=e$ for $g \in \Gamma$ and $n \in {\mathbb N} \Rightarrow g=e$.} discrete group $\Gamma$, subgroup of the isometry group of Euclidean $d$-dimensional space, namely O($d$)$\ltimes {\mathbb R}^d$.
Since we are interested in manifolds allowing for a spin structure, we will constrain ourselves to orientable manifolds and hence focus on $d$-dimensional manifolds $M_d$ defined as
\begin{equation}
		M_d = {\mathbb R}^d/\Gamma, \qquad \Gamma \subset {\rm SO}(d) \ltimes {\mathbb R}^d.
\end{equation}

The elements $\gamma \in \Gamma$ can therefore be described by a rotation matrix and a translation 
\begin{equation}
		\Gamma \ni \gamma = (R, \vec{b}), \qquad R \in {\rm SO}(d), \quad \vec{b} \in {\mathbb R}^d,
\end{equation}
and can be represented by $d+1$-dimensional matrices
\begin{equation}
		\gamma = \left(\begin{array}{cc}
		R & \vec{b} \\
		0 & 1
		\end{array}\right).
\end{equation}

These same manifolds can be viewed as a quotient of a $d$-dimensional torus ${\mathbb T}^d$ in the following way.
The unique Abelian maximal normal subgroup of $\Gamma$ gives a lattice in ${\mathbb R}^d$
\begin{equation}
		\Lambda = \Gamma \cap {\mathbb R}^d,
\end{equation}
and their \emph{crystallographic point group} is given by $r(\Gamma)$, where $r$ is the projection map 
\begin{equation}
		r : {\rm O}(d) \ltimes {\mathbb R}^d \to {\rm O}(d).
\end{equation}
This implies that the lattice is generated by elements in $\Gamma$ of the form $({\mathbb 1},\vec{b})$.
We will call $e_a \in \Lambda$, $a=1,\ldots,d$, the lattice basis and $\vec{e}_a$ their translational part.
In this basis \cite{Charlap1986} all entries of the rotational part of the elements of $\Gamma$, namely $R_{\gamma}$, are given by integer coefficients, though in this basis the matrices are not anymore rotations.
It is then obvious that $\Lambda$ is stable under $r(\Gamma)$ and that $r(\Gamma)$ is the holonomy of the manifold $M_d$.
It is also clear that the d-dimensional torus ${\mathbb T}^d = {\mathbb R}^d/\Lambda$ provides a covering of $M_d$. 
For this reason the fundamental domain of the manifold $M_d$ is generally smaller than the elementary cell defined by the lattice.
The collection of the basis vectors gives an invertible real matrix $A$ defining the lattice $\Lambda = A {\mathbb Z}^d$.

For each lattice $\Lambda$, we introduce the dual lattice
\begin{equation}
		\Lambda^* \equiv \left\{ \vec{k}^* \in {\mathbb R}^d\ |\ \vec{v} \cdot \vec{k}^* \in {\mathbb Z}, \ \forall \vec{v} \in \Lambda \right\}.
\end{equation}
Dual lattice elements can also be written in terms of integers once we introduce the dual basis $\vec{e}_a^{\; *}$, such that $\vec{e}_a^{\; *} \cdot \vec{e}_b = \delta_{ab}$ and
\begin{equation}
		\Lambda^* \ni \vec{k}^* = n_a \vec{e}^{\; *}_a, \qquad n_a \in {\mathbb Z}.
\end{equation}
This same lattice can be described in terms of the transposed inverse of the matrix $A$ above: $\Lambda^* = (A^{-1})^T{\mathbb Z}^d$.

In the following we will often need the norm of the elements of $\Lambda^*$, given by 
\begin{equation}
		\lVert\vec{k}^*\rVert^2 = n_a n_b G_{ab},
\end{equation}
where 
\begin{equation}
	G = A^{-1}(A^{-1})^T
\end{equation}
is the Gram matrix.

The requirement of a spin structure constrains the possible choices of crystallographic groups.
In fact, even if a manifold is orientable, the existence of spinors is not guaranteed.
While in $d=3$ dimensions all orientable manifolds are spin, in $d=4$ only 24 out of 27 orientable manifolds are spin and this counting diverges fast for increasing $d$. 
Currently, complete classifications are available only up to $d=6$, and in this case it is interesting to note that there are $28927922$ inequivalent orbifolds (crystallographic groups): $38746$ of them are flat manifolds (Bieberbach), 3314 are orientable and only 760 admit a spin structure\footnote{$M_d$ has a spin structure if and only if its second Stiefel--Whitney class $w_2(M_d)$ vanishes \cite{Kirby} and if $M_d$ admits at least one spin structure, then all spin
structures correspond to the set $H^1(M_d ; {\mathbb Z}_2)$ \cite{Friedrich}.
} \cite{Cid,Lutowski}.

Spin flat manifolds may actually admit multiple spin structures, and this will be rather important in the following.
We can see how this happens by providing their explicit construction.
Let us start from the Clifford algebra in $d$ dimensions
\begin{equation}
		\{ \gamma^i, \gamma^j\} = 2 \,\delta^{ij},
\end{equation}
where $\gamma^i$ are matrices acting on $\Sigma_d = {\mathbb C}^{2^{[d/2]}}$.
A spin structure requires an epimorphism
\begin{equation}
		\mu : {\rm Spin}(d) \to {\rm SO}(d),
\end{equation}
which provides the double covering of SO($d$) and therefore has ker$\,\mu = \pm {\mathbb 1}$.
We can therefore build different spin structures for a given orientable flat manifold with fundamental group $\Gamma$ by making different choices of homomorphisms
\begin{equation}
		\varepsilon: \Gamma \to {\rm Spin}(d)
\end{equation}
such that 
\begin{equation}
		r = \mu \circ \varepsilon,
\end{equation}
i.e.~the following diagram commutes
\begin{equation}
	\begin{tikzpicture}[baseline=(current  bounding  box.center)]
	\node (g) at (0,2) {$\Gamma$};
	\node (sp) at (3,2) {$\rm{Spin}(d)$};
	\node (so) at (3,0) {$\rm{SO}(d)$}; 
	\draw[->] (g) --node[above left]{$\varepsilon$} (sp);
	\draw[->] (sp) -- node[right]{$\mu$} (so);
	\draw[->] (g) -- node[above]{$r$} (so);
	\end{tikzpicture}\qquad .
\end{equation}
This gives a spin structure and an induced orientation, depending on the $\varepsilon$ map.
In particular, each lattice generator $e_a \in \Lambda$ does not contain any rotational part and hence it can be mapped onto plus or minus the identity in Spin$(d)$:
\begin{equation}
		\varepsilon(e_a) = \delta_a \,{\mathbb 1}, \qquad \delta_a = \pm 1.
\end{equation}
We can therefore have inequivalent spin structures classified by the choices of $\delta_a$. 
The spin structure resulting by the choice $\delta_a = 1$ for all $e_a$ is called trivial.
For instance, the torus ${\mathbb T}^d$ admits $2^d$ different spinor structures, because each $e_a$ can be mapped indifferently to $\pm {\mathbb 1}$.
These are essentially the usual or twisted boundary conditions on fermions according to Scherk--Schwarz \cite{Scherk:1978ta},
\begin{equation}
		\psi(\vec{y} + \vec{e}_a) = \delta_a \, \psi(\vec{y}),
\end{equation}
while generally an equivariant spinor is a map $\psi: M_d \to \Sigma_d$ such that
\begin{equation}
		\psi(\gamma(y)) = \varepsilon(\gamma)\psi(y), \qquad \forall \gamma \in \Gamma.
\end{equation}

While it is clear that different choices of $\delta_a$ provide different spin structures, the consistency constraints of the $\Gamma$ algebra when mapped into Spin$(n)$ may reduce, completely fix or do not allow any consistent choice of $\delta_a$. An algorithmic approach to the problem of the existence of spin structures on flat manifolds is presented in \cite{Lutowski}, where the authors also classified such manifolds in dimension 5 and 6.

\subsection{Examples} 
\label{sub:examples}

Throughout this note, we will focus mostly on two qualitatively different examples of 3-dimensional spin Bieberbach manifolds.
The first one is ${\mathbb T}^3/{\mathbb Z}_3$, with a trivial spin structure, and the second one is the 3-dimensional HW manifold \cite{Handtsche1935}.

\subsubsection{${\mathbb T}^3 / {\mathbb Z}_3$} 
\label{subsubT3}

In the language of this note, the Bieberbach manifold obtained as a ${\mathbb Z}_3$ freely acting orbifold of the 3-dimensional torus is the manifold $M_3 = {\mathbb R}^3/\Gamma$, with $\Gamma$ generated by
\begin{equation}\label{Z3gen1}
		\alpha = \left(\begin{array}{cc}
		R_{\alpha} & \vec{b}_{\alpha} \\
		0 & 1
		\end{array}\right) = \left(
\begin{array}{cccc}
\cos \frac{2 \pi}{3} & \sin \frac{2 \pi}{3} & 0 & 0 \\
 -\sin \frac{2 \pi}{3} & \cos \frac{2 \pi}{3} & 0 & 0 \\
 0 & 0 & 1 & \frac{L_3}{3} \\
 0 & 0 & 0 & 1 \\
\end{array}
\right),
\end{equation}
and 
\begin{equation}\label{Z3gen2}
		e_1 = \left(
\begin{array}{cccc}
 1 & 0 & 0 & L \\
 0 & 1 & 0 & 0 \\
 0 & 0 & 1 & 0 \\
 0 & 0 & 0 & 1 \\
\end{array}
\right), \quad 
		e_2 = \left(
\begin{array}{cccc}
 1 & 0 & 0 & - \frac{L}{2} \\
 0 & 1 & 0 & \frac{\sqrt{3} \, L}{2} \\
 0 & 0 & 1 & 0 \\
 0 & 0 & 0 & 1 \\
\end{array}
\right),
\end{equation} 
where $L_3$ and $L$ are two independent lengths in appropriate units, i.e.
\begin{equation}\label{GammaZ3}
		\Gamma = \langle e_1, e_2, \alpha\ |\ e_1 e_2 = e_2 e_1, \alpha e_1 \alpha^{-1} = e_2^{-1}e_1^{-1},\alpha e_2 \alpha^{-1} = e_1  \rangle.
\end{equation}

It is straightforward to see that the lattice $\Lambda$ is generated by  $e_1$, $e_2$ and $e_3 = \alpha^3$, or 
\begin{equation}
		A = \left(\begin{array}{ccc}
		L & -\frac{L}{2} & 0 \\
		0 & \frac{\sqrt{3} \, L}{2} & 0 \\ 
		0 & 0 & L_3 
		\end{array}\right),
\end{equation}
while the dual lattice is generated by 
\begin{equation}
		e_a^{\; *}=\left(\begin{array}{cc}
		{\mathbb 1} & \vec{e}^{\; *}_a \\
		0 & 1
		\end{array}\right),
\end{equation}
whose translational components are
\begin{equation}\label{dualZ3}
		\vec{e}_1^{\; *} = \left(\begin{array}{c}
		\frac{1}{L}  \\ \frac{1}{\sqrt3 \, L} \\ 0
		\end{array}\right),
		\quad
		\vec{e}_2^{\; *} = \left(\begin{array}{c}
		0 \\ \frac{2}{\sqrt3 \, L} \\ 0
		\end{array}\right),
		\quad
		\vec{e}_3^{\; *} = \left(\begin{array}{c}
		0 \\ 0 \\ \frac{1}{L_3}
		\end{array}\right).
\end{equation}
As expected, in the lattice basis the non-trivial generator $\alpha$ becomes integer valued, because
\begin{equation}
		R_{\alpha}\vec{e}_1 = -\vec{e}_1-\vec{e}_2, \quad R_{\alpha}\vec{e}_2 = \vec{e}_1, \quad R_{\alpha}\vec{e}_3 = \vec{e}_3.
\end{equation}
This implies that $\Lambda$ is invariant under the point group, as is $\Lambda^*$:
\begin{equation}
		R_{\alpha}\vec{e}^{\; *}_1 = -\vec{e}^{\; *}_2, \quad R_{\alpha}\vec{e}^{\; *}_2 = \vec{e}^{\; *}_1-\vec{e}^{\; *}_2, \quad R_{\alpha}\vec{e}^{\; *}_3 = \vec{e}^{\; *}_3.
\end{equation}

We are now ready to introduce a spin structure on this manifold.
The rotational part of the $\alpha$ generator is a rotation by $2\pi/3$ over the third direction and hence, if we introduce as usual the spin generators $J_i = \frac14\, \epsilon_{ijk}\sigma_j \sigma_k = \frac{i}{2} \sigma_i$, we can construct the $\varepsilon$ map by choosing
\begin{equation}  
	\varepsilon(\alpha) = - \delta_3 \exp\left( \frac{2 \pi}{3} \frac12 \sigma_1 \sigma_2\right) = - \delta_3 \left(
\begin{array}{cc}
  e^{\frac{i \pi }{3}} & 0 \\
 0 &  e^{-\frac{i \pi }{3}} \\
\end{array}
\right),
\end{equation}
and
\begin{equation}
	\varepsilon(e_1) = {\mathbb 1}, \quad  \varepsilon(e_2) = {\mathbb 1}, \quad \varepsilon(e_3) = \delta_3 {\mathbb 1},
\end{equation} 
for the lattice generators, where $\delta_3 = \pm 1$.
In fact, possible ambiguities on the signs are fixed by the constraints necessary to fulfill (\ref{GammaZ3}), namely
\begin{equation}
		\varepsilon(\alpha)^3 = \varepsilon(e_3)
\end{equation}
and 
\begin{eqnarray}
		\varepsilon(\alpha)\varepsilon(e_1)\varepsilon(\alpha)^{-1} &=& \varepsilon(e_1)^{-1}\varepsilon(e_2)^{-1}, \\[2mm]
		\varepsilon(\alpha)\varepsilon(e_2)\varepsilon(\alpha)^{-1} &=& \varepsilon(e_1).
\end{eqnarray}
This implies that one can put two spin structures on this manifold, according to the choice of $\delta_3$, and a trivial structure follows from the choice $\delta_3 = 1$.

\subsubsection{Hantzsche--Wendt} 

The HW manifold is a peculiar 3-dimensional manifold that can be obtained by fibering the Klein-bottle over an interval \cite{Handtsche1935}.
As a Bieberbach manifold $M_3 = {\mathbb R}^3/\Gamma$, it can be obtained by a $\Gamma$ generated by 
\begin{equation}
	 	\alpha =  \left(\begin{array}{cccc}
		1 & 0 & 0 & \frac{L_1}{2} \\
		0 & -1 & 0 & 0\\
		0 & 0 & -1 & 0\\
		0 & 0 & 0 & 1
		\end{array}\right), \qquad 
		\beta =\left(\begin{array}{cccc}
		-1 & 0 & 0 & \frac{L_1}{2}\\
		0 & 1 & 0 & \frac{L_2}{2} \\
		0 & 0 & -1 & \frac{L_3}{2} \\
		0 & 0 & 0 & 1
		\end{array}\right),
\end{equation}
\begin{equation}
		\Gamma = \langle \alpha, \beta\ |\ \alpha^2 \beta^2 = \beta^2 \alpha^2, \beta \alpha \beta \alpha^2 = \alpha^2 \beta \alpha \beta, \alpha \beta \alpha \beta^2 = \beta^2 \alpha \beta \alpha \rangle \, ,
\end{equation}
where $L_1$, $L_2$ and $L_3$ are three independent lengths in appropriate units.
The corresponding 3-dimensional lattice follows from the relations
\begin{equation}
		e_1 = ({\mathbb 1}_3,\vec{e}_1) = \alpha^2, \qquad e_2 = ({\mathbb 1}_3,\vec{e}_2) = \beta^2, \qquad e_3 = ({\mathbb 1}_3,\vec{e}_3) = (\beta \alpha)^2,
\end{equation}
and  it is a cubic lattice, as can be seen by the resulting
\begin{equation}
		A = \left(\begin{array}{ccc}
		L_1 & 0 & 0 \\
		0 & L_2 & 0 \\ 
		0 & 0 & L_3 
		\end{array}\right) \, .
\end{equation}
The holonomy group is $r(\Gamma) = {\mathbb Z}_2 \times {\mathbb Z}_2$, and has a diagonal action in this basis:
\begin{equation}
	R_{\alpha} = \left(\begin{array}{ccc}
	1 & 0 & 0 \\
	0 & -1 & 0 \\
	0 & 0 & -1
	\end{array}\right), \qquad 
	R_{\beta} = \left(\begin{array}{ccc}
	-1 & 0 & 0 \\
	0 & 1 & 0 \\
	0 & 0 & -1
	\end{array}\right), \qquad
	R_{\alpha\beta} = R_{\beta \alpha} = \left(\begin{array}{ccc}
	-1 & 0 & 0 \\
	0 & -1 & 0 \\
	0 & 0 & 1
	\end{array}\right).
\end{equation}
In the usual orbifold notation, $M_3$ is the freely acting orbifold defined by the identifications
\begin{equation}
		\left\{\begin{array}{l}
		y^1 \sim y^1 + \frac{L_1}{2} , \\
		y^2 \sim - y^2,\\
		y^3 \sim -y^3,
		\end{array}\right. \quad 
		\left\{\begin{array}{l}
		y^1 \sim - y^1 + \frac{L_1}{2} , \\
		y^2 \sim y^2 + \frac{L_2}{2} ,\\
		y^3 \sim -y^3 + \frac{L_3}{2},
		\end{array}\right. \quad
		\left\{\begin{array}{l}
		y^1 \sim -y^1, \\
		y^2 \sim -y^2 + \frac{L_2}{2} ,\\
		y^3 \sim y^3 + \frac{L_3}{2},
		\end{array}\right.
\end{equation}
with a cubic lattice of length $L_i$ for each $y^i$ ($i=1,2,3$).
The fundamental domain has a volume that is $L_1 L_2 L_3 /4$, though:
\begin{equation}
		D = \left] 0,\tfrac{L_1}{2} \right[ \, \times \, \left]0,\tfrac{L_2}{2}\right[  \, \times \, \left]0,\tfrac{L_3}{2}\right[ \;\; \cup \;\; 
		\left]-\tfrac{L_1}{2},0\right[ \,  \times \, \left]-\tfrac{L_2}{2},0\right[ \, \times \, \left]-\tfrac{L_3}{2},0 \right[\,.
\end{equation}

The holonomy group $r(\Gamma)$ is generated by rotations by $\pi$ about one of the three axis.
We can therefore construct four distinct spinor structures, defined by
\begin{equation}
		\varepsilon(\alpha) = s_1 \exp\left(\pi \frac12\sigma_2 \sigma_3\right) = i\, s_1 \sigma_1,
\end{equation}
\begin{equation}
		\varepsilon(\beta) = s_2 \exp\left(\pi \frac12\sigma_3 \sigma_1\right) = i\, s_2 \sigma_2,
\end{equation}
where $s_i = \pm1$.
However, the closure of the group relations $\alpha^2 = e_1$, $\beta^2 = e_2$, $(\beta\alpha)^2 = e_3$ implies $\delta_a = -1$ for all three lattice generators and hence there are no trivial spin structures on this manifold.
In particular, globally defined spinors satisfy
\begin{eqnarray}
		\psi\left(y^1+\frac{L_1}{2}, -y^2, -y^3\right) &=& i s_1 \,\sigma_1 \psi(y^1,y^2,y^3), \\[2mm]
		\psi\left(-y^1 + \frac{L_1}{2}, y^2+\frac{L_2}{2}, -y^3  + \frac{L_3}{2} \right) &=& i s_2\, \sigma_2 \psi(y^1,y^2,y^3), \\[2mm]
		\psi\left(-y^1, -y^2 + \frac{L_2}{2}, y^3+\frac{L_3}{2} \right) &=& -i s_1 s_2\, \sigma_3 \psi(y^1,y^2,y^3), \\[2mm]
		\psi(y^1+L_1,y^2,y^3) &=& - \psi(y^1,y^2,y^3),\\[2mm]
		\psi(y^1,y^2+L_2,y^3) &=& - \psi(y^1,y^2,y^3), \\[2mm]
		\psi(y^1,y^2,y^3+L_3) &=& - \psi(y^1,y^2,y^3),
\end{eqnarray}
hence showing twisted boundary conditions in all three directions, for any choice of $s_1 = \pm 1$ and $s_2 = \pm 1$.



\section{Harmonics and KK spectrum} 
\label{sec:kaluza_klein_spectrum_on_spin_bieberbach_manifolds}

Since our goal is to derive the full KK spectrum on spin Bieberbach manifolds, we need to understand how to construct globally defined functions, spinors and tensors that provide eigenstates of the Laplace--Beltrami operators used to determine the spectrum.

The Riemann flat manifolds we are discussing are quotients of ${\mathbb R}^d$, hence we can define the tangent frame by means of the coordinate basis and the mass operators of ordinary flat space.
Explicitly, to construct globally defined quantities over $M_d$, we fix a Lorentz gauge so that the vielbeins are identified with the differentials $dy^i$ and we require compensating Lorentz transformation to globally define them, as well as the induced spinorial structure.
With this choice, the Laplace--Beltrami operators relevant for the computation of the KK spectrum are the following:
\begin{equation}\label{box}
	\boxtimes Y_0 = \Box Y_0 = \partial^i \partial_i Y_0 =  \delta^{ij}\partial_i \partial_j Y_0,
\end{equation}
for the scalar harmonic,
\begin{equation}
		\boxtimes Y_{i_1\ldots i_p} = (p+1)\partial^j \partial_{[j} Y_{i_1\ldots i_p]},
\end{equation}
for the p-forms, up to $p = (d-1)/2$, for $d$ odd, where we will use
\begin{equation}
		\boxtimes Y_{i_1 \ldots i_{p}} = \epsilon_{i_1\ldots i_{p}}{}^{j_1 \ldots j_{p+1}}\partial_{j_1} Y_{j_2 \ldots j_{p+1}},
\end{equation}
then
\begin{equation}
		\boxtimes Y_{ij} = 3\partial^k \partial_{(k} Y_{ij)},
\end{equation}
for the symmetric tensor $Y_{ij}= Y_{ji}$ and 
\begin{equation}
		\boxtimes \Xi = \gamma^i \partial_i \Xi,
\end{equation}
for the spin 1/2 and
\begin{equation}\label{spin32}
		\boxtimes \Xi_{i} = \gamma{}_i{}^{jk} \partial_j \Xi_{k},
\end{equation}
for the spin 3/2 harmonics.

We should point out that there are various examples of explicit constructions of the scalar and of some tensor harmonics for these spaces or for some special subset in \cite{Riazuelo:2003ud,Andriot:2018tmb,Peng:2019xoj,Kehagias:2004gy}, while for the spin 1/2 harmonics the prescription for their construction is spelled out in \cite{Pfaffle}. 
Moreover, several statements about the spectrum of Laplace--Beltrami operators on Bieberbach manifolds follow even without an explicit construction \cite{Miatello2001,Miatello2003}.
However, in the following we will provide a general explicit construction of the harmonics, which can be useful for different physical applications.

As we discussed above, spin Bieberbach manifolds can be described as quotients of a torus ${\mathbb T}^d/r(\Gamma)$.
This implies that the harmonics will be obtained from linear combinations of the harmonics on ${\mathbb T}^d$ that are invariant under $r(\Gamma)$.

\subsection{Bosonic harmonics} 
\label{sub:bosonic_harmonics}

Let us start with the scalar harmonics.
All combinations of the form
\begin{equation}\label{scalarharmonic}
		Y_{{\vec k}^*} =\frac{1}{|r(\Gamma)|} \sum_{\gamma \in \Gamma/\Lambda} \exp \left[2 \pi i \, \vec{k}^* \cdot \left(R_{\gamma}\vec{y} + \vec{b}_{\gamma}\right)\right], \qquad \vec{k}^* = n_a \vec{e}^{\; *}_a \in \Lambda^*,
\end{equation}
where $| r(\Gamma)|$ is the order of the holonomy group, are invariant under the action of the holonomy group generators and are also invariant under the lattice generators because $\Lambda$ is $r(\Gamma)$ invariant and $\vec{k}^* \cdot \vec{e_a} \in {\mathbb Z}$.
Such functions are indeed harmonic functions of the scalar Laplacian
\begin{equation}
		\Box Y_{{\vec k}^*}  = - 4 \pi^2 \lVert\vec{k}^*\rVert^2 \,Y_{{\vec k}^*}
		\equiv - M_{\vec{k}^*}^2 \,Y_{{\vec k}^*} \, .
\end{equation}
We can then see that such harmonics are classified by $d$ integers $n_a$, labeling the points of the dual lattice $\Lambda^*$ and the invariant states are going to be generically $1/|r(\Gamma)|$ of those of the covering d-dimensional torus.

We can now see that qualitatively we can split Riemann flat manifolds into two different classes, according to the action of the holonomy group on the coordinates.
We can either have manifolds where no directions are fixed under the action of the holonomy group or we can have manifolds where some directions remain fixed under the same action.
The two examples introduced in the previous section are exactly instances in different classes, with the HW manifold that has no fixed directions and ${\mathbb T}^3/{\mathbb Z}_3$  that has one fixed direction.
Let us look at this last case.
The scalar harmonic following from the definition (\ref{scalarharmonic}) is
\begin{eqnarray}
Y_{n_1,n_2,n_3} & = & \frac{1}{|r(\Gamma)|} \sum_{\gamma \in \Gamma/\Lambda} \exp \left[2 \pi i \, \vec{k}^* \cdot \left(R_{\gamma}\vec{y} + \vec{b}_{\gamma}\right)\right] \nonumber \\
& = & \frac13\sum_{\gamma \in \{e_3, \alpha,\alpha^2\}} \exp \left[2 \pi i \, \vec{k}^* \cdot \left(R_{\gamma}\vec{y} + \vec{b}_{\gamma}\right)\right] \nonumber \\
&= & \frac13\left[
e^{2 \pi i\left(n_1 \frac{y^1}{L} + \frac{n_1 + 2n_2}{\sqrt3} \frac{y^2}{L} + n_3 \frac{y^3}{L_3} \right)} 
+  
e^{2 \pi i\left(-(n_1+n_2) \frac{y^1}{L} + \frac{n_1 - n_2}{\sqrt3} \frac{y^2}{L} + n_3 \left(\frac{y^3}{L_3} +\frac{1}{3}\right) \right)}
\right.
\nonumber \\
&  & \phantom{\frac13 \left[ \right.} + \left.  e^{2 \pi i\left(n_2 \frac{y^1}{L} - \frac{2n_1 + n_2}{\sqrt3} \frac{y^2}{L} + n_3 \left(\frac{y^3}{L_3} +\frac{2}{3}\right)\right)}\right] \, ,
\end{eqnarray}
where
\begin{equation}
	\vec k^* = n_1 \vec{e}_1^{\; *} + n_2 \vec{e}_2^{\; *} + n_3 \vec{e}_3^{\; *} = \left(\begin{array}{c}
		\frac{n_1}{L} \\ \frac{n_1+2 n_2}{\sqrt3 \, L} \\ \frac{n_3}{L_3}
		\end{array}\right).
\end{equation}
These are eigenstates of the scalar Laplacian with eigenvalue $M_{\vec{k}^*}^2 = 4 \pi^2   \lVert\vec{k}^*\rVert^2$ fixed by the norm $\lVert\vec{k}^*\rVert^2 = (4/3) (n_1^2 + n_2^2 + n_1 n_2)/L^2 + n_3^2/L_3^2$.
For generic values $(n_1,n_2,n_3)$, the spectrum multiplicity is 1/3 of the one of the corresponding covering torus, because for each choice of $(n_1,n_2,n_3)$ there are two more choices leading to the same identical harmonic and eigenvalue.
This means that inequivalent harmonics are defined by
\begin{equation}
	n_1,n_2,n_3 \in {\mathbb Z},\quad  n_1 \geq 1, \quad n_2 > -n_1.
\end{equation}
However, we have some special values that correspond to the fixed directions under the holonomy group, namely $R_{\gamma} \, \vec k^* = \vec k^*$.
These are the vectors with labels $n_1=n_2=0$ and arbitrary $n_3$.
Also in this case we see that the allowed harmonics are 1/3 of those of the torus, because the only invariant harmonics are those with $n_3 \in 3 {\mathbb Z}$.
In fact, the invariant construction above gives
\begin{equation}
		Y_{0,0,n_3} = \frac13\,e^{2 \pi i\, n_3\, \frac{y^3}{L_3}}\left(1 + e^{2\pi i\, \frac{n_3}{3} } + e^{2\pi i\, \frac{2n_3}{3} } \right) 
\end{equation}
which is non vanishing for $n_3 \in 3 {\mathbb Z}$ and 0 otherwise.

We can then easily move to the construction of globally defined harmonics for $p$-tensor, by using the same construction, namely
\begin{equation}\label{tensorharmonic}
		Y_{(p)\vec k^*} = \sum_{\gamma \in \Gamma/\Lambda} (R_{\gamma} d\vec{y})^{i_1} \otimes \ldots (R_{\gamma} d\vec{y})^{i_p} \, c_{i_1\ldots i_p} \exp \left[2 \pi i \, \vec{k}^* \cdot \left(R_{\gamma}\vec{y} + \vec{b}_{\gamma}\right)\right],
\end{equation}
for $c_{i_1\ldots i_p}$ are constant coefficients that will be fixed by the corresponding Laplace--Beltrami operator so that 
\begin{equation}
		\boxtimes Y_{(p)\vec k^*} = - 4 \pi^2 \lVert\vec{k}^*\rVert^2 \, Y_{(p)\vec k^*}
		\equiv - M_{\vec{k}^*}^2 \, Y_{(p)\vec k^*} \, .
\end{equation}
At this stage this includes both transverse and longitudinal harmonics (namely harmonics that are the differential of lower-rank tensor harmonics), though for the KK spectrum we will be interested only in the transverse ones.

Also in this case the multiplicity of states with the same eigenvalue $M_{\vec{k}^*}^2$ is reduced with respect to that of the covering torus and the generic formula was provided in \cite{Miatello2001}:
\begin{equation}\label{bosonmultiplicity}
		d_{M_{\vec{k}^*}^2} = \frac{1}{|r(\Gamma)|} \sum_{\gamma \in \Gamma/\Lambda} tr_p(R_{\gamma}) \sum_{\vec{k}^* \in \Lambda^* | R_{\gamma}\vec{k}^* = \vec{k}^*} e^{2 \pi i \vec{k}^* \cdot \vec{b_{\gamma}}},
\end{equation}
where $tr_p$ is the trace of the canonical representation of SO($d$) on the p-forms.
It should be noted that at this stage this multiplicity counts all eigenstates.

Again, for ${\mathbb T}^3 / {\mathbb Z}_3$ we have special series for $n_1=n_2=0$ and $n_3 \neq 0$.
For instance, we have longitudinal 1-forms
\begin{equation}
	\exp\left(2 \pi i n_3 \,\frac{y^3}{L_3}\right) dy^3 , \qquad {\rm{for}} \quad  n_3 \in 3 {\mathbb Z},
\end{equation}
and transverse harmonic 1-forms
\begin{equation}\label{harm1}
	\exp\left(2 \pi i n_3 \, \frac{y^3}{L_3} \right)(dy^1 + i\, dy^2) , \qquad {\rm{for}} \quad n_3 \in 3 {\mathbb Z}+1, \ {\rm or} \ n_3 \in 3 {\mathbb Z}+2.
\end{equation}


\subsection{Fermionic harmonics} 
\label{sub:fermionic_harmonics}

Since we established above that on Bieberbach spin manifolds we can introduce different spinorial structures according to the choice of the homomorphism $\varepsilon$, we should provide harmonics compatible with each different choice.
In particular, we have seen that fermions should be periodic or anti-periodic on the lattice, according to the signs specified by $\varepsilon(e_a) = \delta_a {\mathbb 1}$.
This means that first of all one should modify the dual lattice to take into account the twists $\delta_a$:
\begin{equation} \label{twist}
		\Lambda^*_{\varepsilon} = \Lambda^* + \vec{a}^*_{\varepsilon}, 
\end{equation}
where
\begin{equation}
		e^{2 \pi i \vec{a}^*_{\varepsilon} \cdot \vec{e}_a} = \delta_a.
\end{equation}
In other words the scalar product between elements of $\Lambda$ and $\Lambda^*_{\varepsilon}$ is ${\mathbb Z}$ or ${\mathbb Z} + \frac12$, according to the direction in $\Lambda$ and the choice of $\delta_a$.

Invariant Dirac spinors then have the following expression:
\begin{equation}
		\Xi_{\vec k^*_{\varepsilon}} = \sum_{\gamma \in \Gamma/\Lambda} \exp \left[2 \pi i \, \vec{k}^*_{\varepsilon} \cdot \left(R_{\gamma}\vec{y} + \vec{b}_{\gamma}\right)\right] [\varepsilon(\gamma)]^{-1}\psi_0, \qquad \vec{k}^*_{\varepsilon} \in \Lambda^*_{\varepsilon},
\end{equation}
where $\psi_0$ is a constant spinor and $\vec{k}^*_{\varepsilon} \in \Lambda^*_{\varepsilon}$ implies that 
\begin{equation}
		\vec{k}^*_{\varepsilon} = \vec{k}^* + \vec{a}^*_{\varepsilon} = n_{\varepsilon a} \vec{e}_a^{\; *} = \left(n_a + \frac{1}{4}(1-\delta_a) \right) \vec{e}_a^{\; *} \, , 
\end{equation}
for $n_a \in {\mathbb Z}$.

{}From these spinors we can build the eigenstates of the Dirac operator.
The action of the Dirac operator on $\Xi_{{\vec k}^*_{\varepsilon}}$ gives
\begin{equation}
	\slashed{\partial} \Xi_{\vec k^*_{\varepsilon}} = \sum_{\gamma \in \Gamma/\Lambda} \exp \left[2 \pi i \, \vec{k}^*_{\varepsilon} \cdot \left(R_{\gamma}\vec{y} + \vec{b}_{\gamma}\right)\right] \left(2 \pi i\,  n_{\varepsilon a} \vec{e}_{ai}^{\; *} R^i{}_j\gamma^j \right)[\varepsilon(\gamma)]^{-1}\psi_0, 
\end{equation}
which, using
\begin{equation}
	[\varepsilon(\gamma)]^{-1} \gamma^i [\varepsilon(\gamma)] = R^{i}{}_j \gamma^j,
\end{equation}
reduces to
\begin{equation}
	\slashed{\partial} \Xi_{\vec k^*_{\varepsilon}} = \sum_{\gamma \in \Gamma/\Lambda} \exp \left[2 \pi i \, \vec{k}^*_{\varepsilon} \cdot \left(R_{\gamma}\vec{y} + \vec{b}_{\gamma}\right)\right][\varepsilon(\gamma)]^{-1} \left(2 \pi i\,  n_{\varepsilon a} \vec{e}_{ai}^{\; *} \gamma^i  \right)\psi_0.
\end{equation}
This therefore means that the eigenstates of the Dirac operator follow by introducing
\begin{equation}\label{Xipm}
	\Xi_{\vec k^*_{\varepsilon},\pm} = \sum_{\gamma \in \Gamma/\Lambda} \exp \left[2 \pi i \, \vec{k}^*_{\varepsilon} \cdot \left(R_{\gamma}\vec{y} + \vec{b}_{\gamma}\right)\right] [\varepsilon(\gamma)]^{-1}\left(1 \pm \frac{n_{\varepsilon a} \vec{e}_{ai}^{\; *}}{\lVert\vec{k}^*_{\varepsilon}\rVert} \gamma^i\right)\psi_0,
\end{equation}
so that
\begin{eqnarray}
	\slashed{\partial} \Xi_{\vec k^*_{\varepsilon},\pm} &=& \sum_{\gamma \in \Gamma/\Lambda} \exp \left[2 \pi i \, \vec{k}^*_{\varepsilon} \cdot \left(R_{\gamma}\vec{y} + \vec{b}_{\gamma}\right)\right] \left(2 \pi i\,  n_{\varepsilon b} \vec{e}_{bj}^{\; *} R^j{}_{k}\gamma^k \right)[\varepsilon(\gamma)]^{-1}\left(1 \pm  \frac{n_{\varepsilon a} \vec{e}_{ai}^{\; *}}{\lVert\vec{k}^*_{\varepsilon}\rVert}\gamma^i\right)\psi_0 \nonumber \\[2mm]
	&=&\sum_{\gamma \in \Gamma/\Lambda} \exp \left[2 \pi i \, \vec{k}^*_{\varepsilon} \cdot \left(R_{\gamma}\vec{y} + \vec{b}_{\gamma}\right)\right] [\varepsilon(\gamma)]^{-1}\left(2 \pi i\,  n_{\varepsilon b} \vec{e}_{bj}^{\; *} \gamma^j  \right)\left(1 \pm  \frac{n_{\varepsilon a}  \vec{e}_{ai}^{\; *}}{\lVert\vec{k}^*_{\varepsilon}\rVert}\gamma^i \right)\psi_0 \nonumber\\[2mm]
	&=& \pm 2 \pi \,i\, \lVert\vec{k}^*_{\varepsilon}\rVert\ \, \Xi_{{\vec k}^*_{\varepsilon},\pm} \, . 
\end{eqnarray}
Again, a general formula for the multiplicities of such eigenstates is given in \cite{Miatello2003}.
In particular, for a generic $\vec{k}^*_{\varepsilon}$ we obtain a symmetric spectrum.
On the other hand, as in the case of the bosonic harmonics, special series appear when one has invariant directions under the action of the holonomy group.
In this case, the spectrum of the Dirac operator is asymmetric \cite{Pfaffle}.

For the spin 3/2 operator we can obtain harmonics by using the results of the spin 1/2 representation.
In fact a transverse traceless invariant  spin 3/2 is 
\begin{equation}
		\Xi^{3/2}_{\vec k^*_{\varepsilon}} = \sum_{\gamma \in \Gamma/\Lambda} R_{\gamma}{}^i{}_j dx^{j} c_i\, \exp \left[2 \pi i \, \vec{k}^*_{\varepsilon} \cdot \left(R_{\gamma}\vec{y} + \vec{b}_{\gamma}\right)\right] [\varepsilon(\gamma)]^{-1}\psi_0, \qquad \vec{k}^*_{\varepsilon} \in \Lambda^*_{\varepsilon},
\end{equation}
where the constants $c_i$ are constrained by the requirements that $\gamma^i \Xi_i = 0$ and $\partial^i \Xi_i = 0$.
Once such constraints are imposed, the equation of motion is the same as the Dirac one and therefore we can obtain the corresponding eigenstates by taking the $\pm$ combinations, in perfect analogy with (\ref{Xipm}).

In the case of supergravity compactifications an important question is whether there are Killing spinors, which in this framework is equivalent to having a non-trivial kernel of the Dirac operator.
A general answer is given in \cite{Pfaffle}, where an explicit formula for the dimension of the kernel of the Dirac operator is given by
\begin{equation}
	{\rm dim}({\rm ker} \slashed{\partial}) = \frac{1}{|r(\Gamma)|} \sum_{R\in r(\Gamma)} \chi(R),
\end{equation}
where $\chi(R)$ is the character of the representation.



\section{1-loop potential and examples} 
\label{sec:1_loop_potential_in_the_compactification_of_type_aia_supergravity_to_7d_on_flat_manifolds}

We now discuss the $D$-dimensional 1-loop effective potential  $V_1$ in KK compactifications on Riemann flat spin manifolds.
First we will derive a general form for $V_1$ in a generic compactification from $D+d$ to $D$ dimensions.
Then, after two simple checks on the 1-dimensional orbifold $S^1/{\mathbb Z}_2$,  we will work out in detail the compactification of Type IIA supergravity on the two $d=3$ Bieberbach manifold introduced in Section~2, and discuss a number of interesting properties of the KK spectrum and of the resulting $V_1$.

\subsection{The effective 1-loop potential} 
\label{sub:the_effective_potential}

The 1-loop effective potential in D dimensions is
\begin{equation}
		V_1 = \frac12 \int \frac{d^Dp}{(2\pi)^D} \sum_{I}  (-1)^{F_I} N_I \sum_{\vec k_I^*}\log\left(p^2 + { M}_{\vec{k}_I^*}^2\right),
\end{equation}
where $I$ is a label for the different fields in the spectrum, $F_I$ is the fermion number and $N_I$ counts the associated degrees of freedom.
We should stress that \emph{the sum is performed only on the independent values of $\vec{k}_I^*$ for the harmonics associated to the field $I$ and belonging to the corresponding dual lattice $\Lambda_I^*$ (or shifted lattice $\Lambda_{I \varepsilon}^*$) of momenta in the extra compact $d$ dimensions}. 
In order to obtain a finite result, we need to regularize this sum and we do so by using the zeta function approach (as, for instance, in \cite{Ponton:2001hq}).
This means that we rewrite the 1-loop potential as a derivative 
\begin{equation}
		V_1 = - \frac{d}{ds}\left[\frac12 \int \frac{d^Dp}{(2\pi)^D} \sum_{I} \sum_{\vec k_I^*}(-1)^{F_I}N_I \left(p^2 + { M}_{\vec{k}_I^*}^2\right)^{-s}\right]_{s=0},
\end{equation}
which, after the external momenta integration (under suitable conditions), reduces to 
\begin{equation}
\label{V1Gamma}
		V_1 =  - \frac{1}{2^{D+1} \pi^{D/2} }\,\sum_{I} \sum_{\vec k_I^*}(-1)^{F_I}N_I\frac{d}{ds} \left[ \frac{\Gamma\left(s-\frac{D}{2}\right)}{\Gamma(s)}{M}_{\vec{k}_I^*}^{D-2s}\right]_{s=0}.
\end{equation}
Using that $M_{\vec{k}_I^*} = 2 \pi \lVert \vec{k}_I^* \rVert$ and that each harmonic is labelled by an element of the dual lattice (with the appropriate shifts for the fermionic harmonics with twisted conditions), we can rewrite the terms in the sum above in a closed expression by using the Epstein zeta function \cite{epsteinzeta}, possibly with an overall factor removing redundancies in the series.
This function, for a given $d$-dimensional lattice $\Lambda$, is defined as
\begin{equation}\label{epsteinzetadef}
		Z_{\Lambda}[s,\vec{a},\vec{b}^*] = \sum_{\vec z \in \Lambda} \frac{e^{-2 \pi i \vec{z}\cdot\vec{b}^*}} {\lVert\vec{z} + \vec{a}\rVert^s} \, ,
\end{equation}
where $\vec{z} = z_a \vec{e}_a \in \Lambda$ (i.e. $z_a \in {\mathbb Z}$), $\vec{a} = a_a \vec{e}_a$, $\vec{b}^* = b_a \vec{e}_a^*$, with $a_a, b_a \in {\mathbb R}$, for $\vec{e}_a$ and $\vec{e}_a^{\, *}$ basis for $\Lambda$ and $\Lambda^*$ respectively, and the sum skips $\vec{z} = -\vec{a}$ (see Appendix~A for its properties).
This function generalizes the Riemann and Hurwitz zeta functions to arbitrary $d$ and arbitrary shifts and phases.
For instance, when $d=1$ and we take a unit lattice, we get
\begin{equation}
		Z_{\mathbb Z}[s,a,0] = 2 \, \zeta(s,a).
\end{equation}

The 1-loop expression can therefore be rewritten as
\begin{equation}\label{1loopint}
		V_1 =  - \frac{\pi^{D/2}}{2}\,\sum_{I} (-1)^{F_I}N_I\frac{d}{ds} \left[ \frac{\Gamma\left(s-\frac{D}{2}\right)}{ (2 \pi)^{2s}\Gamma(s) |r(\Gamma_I)|}Z_{\Lambda^*_I}(-D+2s,\vec{a}_I^{\, *},0)\right]_{s=0},
\end{equation}
where $\vec{a}_I^{\, *}$ are the shifts associated to the possible twisted conditions of the fermions, as discussed in (\ref{twist}), and $|r(\Gamma_I)|$ is the factor associated to the redundancy of the harmonics for the expansion of the $I$ field.

An interesting relation that allows us to simplify the computation of the derivative is a functional equation that generalizes the one satisfied by Riemann zeta function \cite{epsteinzeta}:
\begin{equation}\label{functional}
e^{-2\pi i \vec{a}^*\cdot\vec{b}} \frac{\Gamma\left(\frac{d-x}{2}\right)}{\pi^{\frac{d-x}{2}}} Z_{\Lambda^*}[d-x,\vec{a}^*,-\vec{b}] = 	\sqrt{\Delta}\,		\frac{\Gamma(x/2)}{\pi^{x/2}} Z_{\Lambda}[x,\vec{b},\vec{a}^*] ,
\end{equation}
where $\Delta$ is the determinant of the matrix defining the norm over the lattice $\Lambda$ and $d$ is the dimension of the lattice.

By adapting (\ref{functional}) to the one in (\ref{1loopint}), using $x = d-2s+D$, we reduce the computation of $V_1$ to a sum over the lattice of the Bieberbach manifold
\begin{equation}
		V_1 =  - \frac{\sqrt{\Delta}}{2 \pi^{\frac{D+d}{2}} }\,\sum_{I} (-1)^{F_I}N_I\frac{d}{ds} \left[ \frac{\Gamma\left(\frac{D+d}{2}-s\right)}{2^{2s}\Gamma(s)|r(\Gamma_I)|}Z_{\Lambda_I}(D+d-2s,0,\vec{a}_I^*)\right]_{s=0},
\end{equation}
and using the fact that the Euler Gamma function diverges in zero, and that for the derivative of its inverse we get  $\left[ d\Gamma^{-1}(s)/ds \right]_{s=0} = 1$, we end up with two expressions
\begin{equation}\label{final1loop}
	\begin{split}
				V_1 &= - \frac{\sqrt{\Delta}\Gamma\left(\frac{D+d}{2}\right)}{2 \pi^{ \frac{D+d}{2}} }\,\sum_{I} (-1)^{F_I}\, \frac{N_I}{|r(\Gamma_I)|}\, Z_{\Lambda_I}(D+d,0,\vec{a}_I^*) \\[2mm]
				&= - \frac{\pi^{D/2}}{2} \, \Gamma(-D/2)\,\sum_{I} (-1)^{F_I}\, \frac{N_I}{|r(\Gamma_I)|}\, Z_{\Lambda_I^*}(-D,\vec{a}_I^*,0),
	\end{split}
\end{equation}
equivalent upon using (\ref{functional}).

It is useful to point out that $V_1$ in (\ref{V1Gamma}) can also be rewritten in terms of supertraces
\begin{equation}
 {\rm Str} \, {\cal M}^{2p} 
 \equiv
 \sum_I \, 
(-1)^{F_I} \,
N_I \,
{M}_{\vec{k}_I^*}^{2p}
\, .
\label{strdef}
\end{equation}
as
\begin{equation}\label{1loopStr}
		V_1 =  - \frac{1}{2^{D+1} \pi^{D/2} }\,\sum_{\vec k^*}\frac{d}{ds} \left[ \frac{\Gamma\left(s-\frac{D}{2}\right)}{\Gamma(s)}{\rm Str} {\cal M}^{D-2s}\right]_{s=0},
\end{equation}
where ${\rm Str} {\cal M}^{2p}$ is a function of the labels of the harmonics $\vec{k}^*$,
and this explains the usual 1-loop expression in ordinary supergravity theories in 4-dimensions
\begin{equation}\label{1loopsugra}
		V_{red} = \frac{1}{32 \pi^2} {\rm Str} {\cal M}_0^2 \, \mu^2 + \frac{1}{64 \pi^2} {\rm Str} \left[{\cal M}_0^4 \log \frac{{\cal M}_0^2}{\mu^2}\right],
\end{equation}
where the $0$ in the pedix is to signal that we are summing only over the 4-dimensional supergravity multiplet states and $\mu$ is some ultraviolet cutoff scale.

It was proved in \cite{DallAgata:2012tne} that for Minkowski vacua breaking supersymmetry in $N=8$ gauged supergravity such potential is finite because there are cancellations in the supertraces that remove the divergent terms in (\ref{1loopsugra}).
As we will see momentarily, we find that there are actually analogous cancellations for  certain specific powers $2n$ of ${\rm Str}\,{\cal M}^{2n}$, before taking the sum over the (dual) lattice.

We stress that, whatever the number $N > 0$ of supersymmetries in the higher dimensional theory, regularized sum over {\em all} KK levels always gives a finite $D=4$ 1-loop effective potential, irrespectively of the presence of modular invariance and of a string cutoff~\footnote{Different supertrace relations and expressions of the one-loop vacuum amplitude have been derived for string constructions with broken supersymmetry, emphasizing the role of modular invariance and the contributions of massive string states~\cite{Dienes:1995pm,Abel:2015oxa,Abel:2024twz}.}.     
This is a consequence of the non-local nature of supersymmetry breaking and is technically explained by the property of the Epstein zeta function $Z_{\Lambda}[-2n,\vec{a},\vec{b}^*] = 0$ ($\forall n\in {\mathbb N}$), which reduces to the well known property of the zeta function $\zeta(-2n)=0$  ($\forall n\in {\mathbb N}$) for a single tower. 
Therefore $V_1$ is finite even if  ${\rm Str} \, {\cal M}^{2n} \ne 0$ at each KK level, since the sum over all the KK spectrum is anyway vanishing. 
 

\subsection{Examples} 
\label{sub:some_examples}

As we have seen above, $V_1$ depends crucially on the KK spectrum.
Since the harmonics on the internal manifolds are specified by their lattices, we can in principle apply formula (\ref{final1loop}) to all spin Riemann flat manifolds, for different choices of $d$ and $D$.
However, since in this paper our interest lies in the presentation of some remarkable identities and properties of these manifolds rather than in a full classification of the possible results, we leave the latter for future work.
Indeed, the explicit form of all generators in the integer basis is provided in \cite{Lutclass} and can be used to construct all KK spectra and compute $V_1$.

We therefore illustrate the computation of the KK spectrum and of $V_1$ for some simple instances, for which we either have a comparison with known results or which have been discussed in the previous sections, leading to a clear illustration of the formalism.

\subsubsection{1-dimensional orbifolds} 
\label{sub:1-dimensional_orbifolds}

The first, very simple instances we can focus on are cases with $d=1$.
There is not really much of a choice of Riemann-flat manifolds in this case and the computation will not give new results, but it is a useful first check.
We will therefore look at 11-dimensional supergravity on the orbifold $S^1/{\mathbb Z}_2$, for which the computation was performed in \cite{Fabinger:2000jd}, and at pure supergravity in 5 dimensions, again on $S^1/{\mathbb Z}_2$, for which the 1-loop potential has been computed in \cite{Bagger:2001ep}.
In both examples we denote by $L$ the length of the $S^1$ perimeter.

In the first example we have a 1-dimensional sum $d=1$ and the reduced theory has $D=10$.
Clearly the reduction is really on an interval and not a smooth manifold, but we can still argue for the correct results if we neglect states localized at the fixed points.
Also, the reduction scheme requires a shift for the fermions $a_F^*= 1/(2L)$, while $a_B^* = 0$.
The degrees of freedom are $N_B = N_F = 128$, but only half of the harmonics are invariant, i.e.~$|r({\mathbb Z}_2)|= 2$.
Inserting all this information in (\ref{final1loop}) gives
\begin{equation}
	\begin{split}
		V_1 &= - L \,\frac{\Gamma\left(\frac{11}{2}\right)}{2 \pi^{\frac{11}{2}}} \frac{128}{2}\left[Z_{L{\mathbb Z}}(11,0,0)-Z_{L{\mathbb Z}}(11,0,1/(2L))\right] \\[2mm]
		&= - \frac{\pi^{5}}{2}\, 64\,\Gamma(-5)\left[Z_{{\mathbb Z}/L}(-10,0,0)-Z_{{\mathbb Z}/L}(-10,1/(2L),0)\right] \\[2mm]
		& =  - \frac{\pi^{5}}{2L^{10}}\, 64\,\Gamma(-5)(2 \zeta(-10) - 2 \zeta(-10,1/2)) \\[2mm]
		&= - \frac{2^{-4}\pi^{5}}{L^{10}}\,(2^{11}  - 1)\,\Gamma(-5)\zeta(-10) = - \frac{\Gamma\left(\frac{11}{2}\right)}{2^4 L^{10} \, \pi^{\frac{11}{2}}}(2^{11}-1)\zeta(11), 
	\end{split}
\end{equation}
which matches the value computed by \cite{Fabinger:2000jd}, upon setting $L = 2 \tilde{L}$, where $\tilde{L}$ is the length of the interval.

{}For the other example we have $D=4$ and $d=1$, again with a shift $a_F^*=1/(2L)$ because of the twisted conditions on the fermions.
We are discussing minimal pure supergravity in 5 dimensions, hence $N_B = N_F = 8$ and again $|r({\mathbb Z}_2)|= 2$.
Hence
\begin{equation}
	\begin{split}
		V_1 &= - L \frac{\Gamma\left(\frac{5}{2}\right)}{2 \pi^{\frac{5}{2}}} \frac{8}{2}\left[Z_{L{\mathbb Z}}(5,0,0)-Z_{L{\mathbb Z}}(5,0,1/(2L))\right] \\[2mm]
		&=  - \frac{\pi^{2}}{2}\, 4\,\Gamma(-2)\left[Z_{{\mathbb Z}/L}(-4,0,0)-Z_{{\mathbb Z}/L}(-4,1/(2L),0)\right]\\[2mm]
		& =  - \frac{\pi^{2}}{2L^{4}}\, 4\,\Gamma(-2)(2\zeta(-4) - 2 \zeta(-4,1/2))= - \frac{\pi^{2}}{4L^{4}}\,(2^{5}  - 1)\,\Gamma(-2)\zeta(-4) \\[2mm]
		&= - \frac{\Gamma\left(\frac{5}{2}\right)}{4 L^{4} \, \pi^{\frac{5}{2}}}(2^{5}-1)\zeta(5) = - \frac{3}{\pi^2 {L}^{4}} \,\frac{31}{16} \,\zeta(5),
	\end{split}
\end{equation}
which coincides with the expression (6.4) in \cite{Bagger:2001ep} once we use $L = 2 \pi R$, and the identity
\begin{equation}
		\zeta(5) - {\rm{Li}}_5(-1) = \frac{31}{16} \zeta(5).
\end{equation}


\subsubsection{Type IIA examples with $d=3$} 
\label{sub:the_kaluza_klein_spectrum}


We now turn to the compactification of Type IIA supergravity on $d=3$ spin Riemann flat manifolds, focussing on the two main examples given in Section~2, namely the ${\mathbb T}^3/{\mathbb Z}_3$ and the HW manifolds.

The computation of the KK spectrum requires a linearization of the Type IIA equations of motion with respect to the fluctuations above the flat background and to do so one has to fix appropriate gauges for the internal diffeomorphisms and 10-dimensional supersymmetries.
After such procedure, the linearized equations of motion have masses specified by the Laplace--Beltrami operators (\ref{box})--(\ref{spin32}).
A simple choice that leads to the expected equations for the fluctuations and allows for a simple harmonic expansion of the 10-dimensional fields is \cite{Duff:1986hr}
\begin{equation}
	\partial^i \left(g_{ij}-\frac13 \delta_{ij}\,g^k{}_k\right) = 0 = \partial^i g_{\mu i} = \partial^i A_i = \partial^i B_{\mu i} = \partial^i B_{ij} = \partial^i C_{\mu ij} = \partial^i C_{ijk},
\end{equation}
for the bosonic fields and \cite{DAuria:1983rmi}
\begin{equation}
		\Gamma^i \psi_i(x,y) = 0,
\end{equation}
for the gravitino.
For the tensor fields, once we employ transverse traceless harmonics for the expansion, we have further simplifications, because of the fact that the internal space is 3-dimensional.
For instance the expansion of the internal 3-form should be given in terms of scalar harmonics:
\begin{equation}
	C_{ijk}(x,y) = \epsilon_{ijk} c^I(x) Y^I(y).
\end{equation}
However, transversality implies $\partial^i C_{ijk}(x,y) = c^I(x) \epsilon_{ijk} \partial^i Y^I(y) = 0$ and hence only the constant scalar harmonic gives a mode in the expansion. 
A similar argument for $C_{\mu ij}$ and $B_{ij}$ implies $\partial^i B_{ij} = b^I(x) \epsilon_{ijk} \partial^i Y_k^I = 0$ and this forces to keep only the closed harmonic 1-forms  plus the closed non-exact 1-forms in the expansion.

For the fermions, we should also take into consideration the split of the 10-dimensional gamma matrices $\Gamma$ in terms of those in $D=7$ and those in $d=3$.
Since a spinor in 10-dimensions has 32 components, a possible split is the following 
\begin{equation}
	\Gamma^\mu = \gamma^\mu \otimes {\mathbb 1}_2 \otimes \sigma_1, \quad \Gamma^i = {\mathbb 1}_8 \otimes \sigma^i \otimes \sigma^2.
\end{equation}
The last matrix acts on the chiral components $\lambda^A$, $A=1,2$, of the 10-dimensional spinor as follows
\begin{equation}
	\lambda_{10} = \left(\begin{array}{c}
	\lambda^1 \\ \lambda^2
	\end{array}\right).
\end{equation}

Overall the various fields expansions are summarized in Table \ref{tabexp}, where  transverse harmonics have been employed (namely $\partial^i Y_i^I = \partial^i Y_{ij}^I = Y_{kk}^I = \sigma^i \Xi_{3/2 i}^I =0$).
\begin{table}
	\begin{center}
{\small
$$	\begin{array}{|cccc|cccc|}\hline
	{\rm field} & spin & {\rm IIA\ origin}  & {\rm harmonic} & 		{\rm field} & spin & {\rm IIA\ origin}  & {\rm harmonic}  \\[2mm]\hline
	h_{\mu\nu}& 2 & g_{\mu\nu} & Y_0    & \psi^A & 1/2 & \psi_i & \Xi_{3/2}\\[2mm]
	\psi_\mu^A & 3/2 & \psi_\mu & \Xi_{1/2}& \eta^A & 1/2 & \psi_i & \Xi_{1/2} \\[2mm]
	a_\mu & 1 & A_\mu & Y_0      & \chi^A & 1/2 & \chi & \Xi_{1/2}\\[2mm]
	h_\mu & 1 & g_{\mu i} & Y_1     &  \phi& 0 & \phi & Y_0 \\[2mm]
	b_\mu & 1 & B_{\mu i} & Y_1     & g & 0 & g_{ij} & Y_{sym} \\[2mm]
	c_\mu & 1 & C_{\mu ij} & \star d Y_0    & h& 0 & g_{ij} & Y_0 \\[2mm]
	b_{\mu\nu}& {\rm tensor} & B_{\mu\nu} & Y_0 & a & 0 & A_i & Y_1 \\[2mm]
	c_{\mu\nu}& {\rm tensor} & C_{\mu\nu i} & Y_1 & b&  0 & B_{ij} & \star dY_0\\[2mm]
	c_{\mu\nu\rho} &{\rm tensor} & C_{\mu\nu \rho} & Y_0 &c &  0 & C_{ijk} & constant 	\\[2mm] \hline
	\end{array}
	$$}
	\end{center}
	\caption{Field expansions in terms of the harmonics.}
\label{tabexp}
\end{table}

Each of the resulting 7-dimensional fields has a number of degrees of freedom that depend on their mass as follows
\begin{equation}
	\begin{array}{|c|c|c|c|c|c|}\hline
	{\rm state} & M \neq0 & M=0 & 	{\rm state} & M\neq0 & M=0 \\[2mm]\hline
	 g_{\mu\nu} & 20 & 14  &	 c_{\mu\nu\rho} & 20 & 10 \\[2mm]
	 \psi_{\mu} & 20 & 16  &	 \chi & 4 & 4 \\[2mm]
	 a_\mu & 6 &5&	 \phi & 1 & 1\\[2mm]
	 b_{\mu\nu} & 15 & 10&&&\\\hline
	\end{array}\,,
\end{equation}
which is compatible with the standard reduction on the straight torus that gives $128$ bosonic and $128$ fermionic states.

Using the information above, we can now compute the full spectrum on ${\mathbb T}^3/{\mathbb Z}_3$ and on the HW manifold.
For  ${\mathbb T}^3/{\mathbb Z}_3$, with trivial fermionic structure, the symmetric part of the spectrum is straightforward, because all states have the same masses in terms of the quantum numbers, because
\begin{equation}
	\lVert\vec{k}^*\rVert^2 = \left[\frac{n_1^2}{L^2} + \frac{(n_1+2n_2)^2}{3L^2} + \frac{n_3^2}{L_3^2} \right], \qquad (\ n_1,n_2,n_3 \in {\mathbb Z},\;  n_1 \geq 1, \; n_2 > -n_1) \, ,
\end{equation}
for all bosons and fermions, and we inserted the dependence on the two size moduli $(L,L_3)$, compatible with the global identifications of the manifold.
Moreover, these states are grouped always in $128$ bosonic and $128$ fermionic states for each $(n_1,n_2,n_3)$, like those of a supersymmetric multiplet, even if supersymmetry is broken, which means that all supertraces contributions from these states exactly cancel level-by-level, as well as their contribution to $V_1$.

The short series on the other hand is asymmetric.
This means that interesting contributions come from the states with $n_1=n_2=0$ and $n_3 \neq 0$.
The harmonics for the short series, for $n_3 \in {\mathbb Z}$, give the following states:
\begin{equation}
	\begin{array}{ccc}\hline
	{\rm harmonic} & {\rm degeneracy} & L_3^2\lVert\vec{k}^*\rVert^2 \\\hline \hline
	Y_0 &1 & 9 n_3^2 \\\hline
	\Xi_{1/2} &2 & (3n_3+1)^2  \\\hline
	Y_1 & 2& (3n_3+1)^2 \\\hline
	\Xi_{3/2} &2 & 9n_3^2   \\\hline
	Y_{sym} & 2 & (3n_3+1)^2  \\\hline
	\end{array}
\end{equation}
Again, for any $n_3\neq 0$, we get the same number of bosonic and fermionic degrees of freedom.
For $n_3=0$ however we have a mismatch, but one should note that in addition to the KK towers above, we get  other massless states associated to the closed non-exact forms. 
The manifold has indeed a closed, non-exact 1-cycle, and this gives additional massless states in the expansion of the harmonics.
We can indeed add a constant harmonic 1-form $Y = dy^3$, resulting in additional massless fields for $g_{\mu i}$, $B_{\mu i}$, $C_{\mu\nu i}$ and $A_i$, its dual $Y_2 = dy^1 \wedge dy^2$, resulting in additional massless states for $C_{\mu ij}$ and $B_{ij}$, a constant symmetric tensor $Y_{33}$, giving one additional scalar in $g_{ij}$, and finally the constant state associated to the volume in $C_{ijk}(x,y)$, which is not part of any particular tower.
At the same time, there is no $n_3=0$ state in the expansion with $\star dY_0$.
Therefore, the $n_3=0$ spectrum, with the additional massless states, contains a massless graviton (14), 4 massive gravitini (80), 4 massless and 4 massive vectors (44), 2 massless and 2 massive 2-forms (50), 1 massless 3-form (10), 12 massive spin 1/2 fields (48) and 6 massless and 4 massive scalars (10).
Hence once more we have 128 bosonic and 128 fermionic degrees of freedom, as should be expected for a spontaneous supersymmetry breaking scenario.
On the other hand the background is fully non supersymmetric and the complete massless spectrum contains 70 massless bosonic degrees of freedom and no fermions.

If we compute $V_1$, the matching of the bosonic and fermionic degrees of freedom removes quartic divergencies even without any regularization.
The other supertraces also satisfy very interesting relations resulting in a finite 1-loop correction even before summing over the KK towers, if we relabel the various states as shown in Table \ref{tabellamasse}.

\begin{table}
	\begin{center}
{\small
$$	\begin{array}{c|c|c}\hline
	{\rm state} & {\rm dofs} & [n_3]-{\rm level} \\[2mm]\hline
	 g_{\mu\nu} & 20 & |3[n_3]|   \\[2mm]
	 \psi_{\mu} & 20 &2 \times |3[n_3]+1|, \ 2 \times |3[-n_3]+1|, \\[2mm]	 
	 a_\mu & 6 &   2 \times |3[n_3]| \oplus 2 \times |3[n_3-1]+1| \oplus 2 \times |3[-n_3-1]+1| \\[2mm]
	 b_{\mu\nu} & 15 & |3[n_3]| \oplus |3[n_3-1]+1| \oplus  |3[-n_3-1]+1| \\[2mm]
	 c_{\mu \nu \rho} &20 & |3[n_3]| \\[2mm]
	 \chi & 4 & 2 \times |3[n_3+1]| \oplus2 \times |3[-n_3+1]| \oplus 4 \times |3[n_3]+1| \oplus 4 \times |3[-n_3]+1| \\[2mm]
	 \phi & 1 & 3 \times |3[n_3]| \oplus |3[-n_3+1]+1| \oplus |3[n_3-1]+1| \oplus |3[-n_3-1]+1| \oplus |3[n_3+1]+1| \\\hline
	\end{array}
	$$}
	\end{center}
	\caption{The shifts in the KK levels, represented by the integer number in square brackets,  for the short series of harmonics for ${\mathbb T}^3/{\mathbb Z}_3$ with trivial spin structure.}
\label{tabellamasse}
\end{table}
With this assignment, for any $n_3$, we get
\begin{equation}
		{\rm Str}({\cal M}^2) = {\rm Str}({\cal M}^4) = {\rm Str}({\cal M}^6) = 0,
\end{equation}
while
\begin{equation}
		{\rm Str}({\cal M}^8) = 40320 \frac{\pi^8}{L_3^8},
\end{equation}
which is non-vanishing and $n_3$-independent.
We therefore see that also in this example all re-arranged KK levels have vanishing supertraces up to the one with power 8, which anyway has a level-independent expression.

These relations also guarantee the finiteness of the contribution to $V_1$ from each redefined KK level,  even before employing the zeta function regularization, because all divergent terms are proportional to the vanishing supertraces above.

We are now in a position to compute the resulting $V_1$ in $D=7$, by using (\ref{1loopStr}) and recalling that the only non-zero contribution comes from the short (asymmetric) spectrum.
Since effectively the only harmonics contributing to $V_1$ depend exclusively on the circle of length $L_3$, the sum can be expressed in terms of the Hurwitz zeta function:
\begin{equation}\label{Z3norm}
	\begin{split}
	V_1 &= - \frac{54 \cdot3^7 \pi^{\frac{7}{2}}}{2 L_3^7} \Gamma\left(-\frac72\right)  \left[2 \zeta(-7)-(\zeta(-7,1/3)+\zeta(-7,2/3))\right]
\\[2mm]
	&= - \frac{162}{\pi^{4}L_3^7}  \left[\zeta(8,1/3)+\zeta(8,2/3)\right]
	= - \frac{162}{\pi^{4}L_3^7}  \left( 3^8 - 1 \right) \, \zeta(8) =- \frac{3936}{35}\frac{\pi^4}{L_3^7} \, ,
	\end{split}
\end{equation}
which agrees with the structure of the leading contribution in the large $L_3$ limit of the string computation in \cite{Acharya:2020hsc}, up to an overall normalization factor.
 
If we had chosen twisted boundary conditions, on the other hand, we would have still obtained a finite $V_1$, once more after relabeling the states in the long and short towers as shown in Table \ref{tabellamassetwist}.
\begin{table}
	\begin{center}
{\footnotesize
$$	
	\begin{array}{c|c|c}\hline
	{\rm state} & {\rm dofs} & (n_1',n_2',n_3')\\[2mm]\hline
	 g_{\mu\nu} & 20 & (n_1,n_2,n_3)   \\[2mm]
	 \psi_{\mu} & 20 &2 \times \left(n_1,n_2,n_3+\frac12\right), \ 2 \times \left(n_1,n_2,n_3-\frac12\right), \\[2mm]	 
	 a_\mu & 6 &   2 \times (n_1,n_2,n_3)\oplus 2 \times \left(n_1,n_2,n_3+1\right)\oplus 2 \times \left(n_1,n_2,n_3-1\right) \\[2mm]
	 b_{\mu\nu} & 15 & (n_1,n_2,n_3) \oplus \left(n_1,n_2,n_3+1\right)\oplus  \left(n_1,n_2,n_3-1\right) \\[2mm]
	 c_{\mu \nu \rho} &20 & (n_1,n_2,n_3) \\[2mm]
	 \chi & 4 & 2 \times \left(n_1,n_2,n_3+\frac12\right) \oplus2 \times \left(n_1,n_2,n_3-\frac12\right) \oplus 4 \times \left(n_1,n_2,n_3+\frac32\right) \oplus 4 \times \left(n_1,n_2,n_3-\frac32\right) \\[2mm]
	 \phi & 1 & 3 \times (n_1,n_2,n_3) \oplus \left(n_1,n_2,n_3+1\right) \oplus \left(n_1,n_2,n_3-1\right) \oplus \left(n_1,n_2,n_3+2\right) \oplus \left(n_1,n_2,n_3-2\right) \\\hline
	\end{array}
	$$}
	\end{center}
	\caption{The shifts in the KK levels for the states of  ${\mathbb T}^3/{\mathbb Z}_3$ with non-trivial spin structure. }
\label{tabellamassetwist}
\end{table}

These identifications give vanishing supertraces up to 
\begin{equation}
		{\rm Str} {\cal M}^8 = 40320\, \frac{\pi^8}{L_3^8},
\end{equation}
which is also $(n_1,n_2,n_3)$-independent.
In this case the 1-loop contribution for the long series is non-vanishing and one can again employ (\ref{final1loop}) to compute its magnitude.
In particular, numerical estimates can be obtained by using the Python library created in \cite{Epsteinzetanum} and in the large $L_3$ limit recover the results in \cite{Acharya:2020hsc}.
In fact the explicit expression is the sum of two contributions, one that comes from the harmonics with generic $n_a \neq 0$ and one from the short series (with $n_1=n_2=0$ and $n_3\neq 0$).
The first one can be written as
\begin{equation}
	\begin{split}
		V_1^{(I)} =&
		 -\frac{\sqrt{3}}{2} {L^2 L_3} \frac{\Gamma(5)}{2 \pi^{5}} \frac{128}{3} \left[ Z_{\Lambda}(10,\vec{0},\vec{0})-Z_{\Lambda}(10,\vec{0},(0,0,1/(2L_3)))\right. \\[2mm]
		 & \left.		-\frac{2}{L_3^{10}}(\zeta(10)+\eta(10))\right] \, ,		
	\end{split}
\end{equation}
where the first line sums all contributions from harmonics on ${\mathbb T}^3$ with the proper complex structure and moduli for the lattice $\Lambda$ described in section \ref{subsubT3}, weighted with a factor $1/3$ to remove the redundancy on the harmonics with $(n_1,n_2)\neq (0,0)$, while the second line removes the harmonics that have only $n_3 \neq 0$. 
The $\eta$-function $\eta(s) = -\sum_{n\in {\mathbb N}} e^{i \pi n}n^{-s}$ accounts for the alternating sum due to the shift in the fermions and satisfies $\eta(s) = (1-2^{1-s})\zeta(s)$.
The second contribution, coming from the short series, is
\begin{equation}
	\begin{split}
		V_1^{(II)} = &- \frac{3^7\pi^{\frac{7}{2}}}{2 L_3^7} \Gamma(-7/2) \left(140 \zeta(-7)+58(\zeta(-7,1/3)+\zeta(-7,2/3)) \right. \\[2mm]
			& \left.- 112
			 (\zeta(-7,1/6)+\zeta(-7,5/6))-32 \zeta(-7,1/2)\right) = -\frac{197}{105} \frac{\pi^4}{L_3^7}.
		\end{split}
\end{equation}
In the large $L_3$ limit the second contribution dominates and indeed we recover the result in \cite{Acharya:2020hsc}, up to the same normalization factor as in (\ref{Z3norm}), so that the ratio of the two expressions in the large $L_3$ limit coincides with the one in \cite{Acharya:2020hsc}.
In fact, if the ratio $L_3/L$ is large, the masses in the sum in the Epstein function $Z_{\Lambda}$ become supersymmetric at the leading order, with corrections of the order of $L/L_3$, hence the leading terms in $V_1^I$ will go like $L^3/L_3^{10}$, which is suppressed with respect to $V_1^{II}$.

Surprisingly, the same procedure can be applied to the HW manifold, which allows only for non-trivial fermionic structures and cannot be related directly to twisted tori reductions, because of the absence of fixed directions under the holonomy group.
A relabeling of the various states appearing in the KK expansion still allows to find vanishing supertraces up to ${\cal M}^8$, for any choice of the fermionic structure and for each single redefined KK level.
In Table \ref{tabellamasseHW} we give here the assignment for the HW manifold.
\begin{table}
	\begin{center}
{\small
$$	
	\begin{array}{c|c|c}\hline
	{\rm state} & {\rm dofs} & (n_1',n_2',n_3')\\[2mm]\hline
	 g_{\mu\nu} & 20 & (n_1,n_2,n_3)   \\[2mm]
	 \psi_{\mu} & 20 &2 \times \left(n_1+\frac12,n_2+\frac12,n_3+\frac12\right), \ 2 \times \left(n_1-\frac12,n_2-\frac12,n_3-\frac12\right), \\[2mm]	 
	 a_\mu & 6 &   2 \times (n_1,n_2,n_3)\ , \ 2 \times \left(n_1+1,n_2+1,n_3+1\right)\ , \ 2 \times \left(n_1-1,n_2-1,n_3-1\right) \\[2mm]
	 b_{\mu\nu} & 15 & (n_1,n_2,n_3) \ , \ \left(n_1+1,n_2+1,n_3+1\right)\ , \  \left(n_1-1,n_2-1,n_3-1\right) \\[2mm]
	 c_{\mu \nu \rho} &20 & (n_1,n_2,n_3) \\[2mm]
	 \chi & 4 & 2 \times \left(n_1+\frac12,n_2+\frac12,n_3+\frac12\right) \ , \ 2 \times \left(n_1-\frac12,n_2-\frac12,n_3-\frac12\right) \ ,  \\[2mm]
	  &  & 4 \times \left(n_1+\frac32,n_2+\frac32,n_3+\frac32\right) \ , \ 4 \times \left(n_1-\frac32,n_2-\frac32,n_3-\frac32\right) \\[2mm]
	 \phi & 1 & 3 \times (n_1,n_2,n_3) \ , \ \left(n_1+1,n_2+1,n_3+1\right) \ , \ \left(n_1-1,n_2-1,n_3-1\right) \ , \\[2mm]
	 &  &	 \left(n_1+2,n_2+2,n_3+2\right) \ , \ \left(n_1-2,n_2-2,n_3-2\right) \\\hline
	\end{array}
	$$}
	\end{center}
	\caption{The shifts in the KK levels for the harmonics on the HW manifold. }
\label{tabellamasseHW}
\end{table}

Since the lattice is straight, the corresponding eigenvalues are
\begin{equation}
		\lVert\vec{k}^*\rVert^2 = \left[\frac{n_1^2}{L_1^2}+\frac{n_2^2}{L_2^2}+\frac{n_3^2}{L_3^2}\right] \\[2mm]
\end{equation}
For such assignment it follows that
\begin{equation}
		{\rm Str} {\cal M}^8 = 40320\, \frac{\pi^8\left(L_1^2 L_2^2 + L_2^2 L_3^2 + L_1^2 L_3^2\right)^4}{L_1^8 L_2^8 L_3^8}
\end{equation}
and the supertraces vanish for lower powers of the masses.
Once  more $V_1$ can be obtained by applying (\ref{final1loop}).
For the choice $L_1 = L_2 = L_3 = L$, this gives
\begin{equation}
		V_1 = - \frac{384}{\pi^{5} L^7}\left( Z_{{\mathbb Z}^3}(10,0,0)-Z_{{\mathbb Z}^3}\left(10,0,\left(-\frac12,-\frac12,-\frac12\right) \right)
		-6\left(\zeta(10)+\eta(10)\right)\right),
\end{equation}
where we moved all scales outside the Epstein zeta function evaluation.
Since the only surviving harmonics have at least two $n_a \neq 0$, we remove from the first sum the terms with a single non-vanishing $n_a$ and divide by a factor 4 for the redundancy due to the freely acting orbifold.

The procedure above shows that we can actually obtain the same supertrace properties for all Bieberbach manifolds, as we will now explain, by appropriately taking into account possible shifts in the values of the integers defining the states in a given level.


\subsection{General supertrace formulae} 
\label{sub:general_supertrace_formulae}

Now that we have understood how to compute the full spectrum for compactifications of maximal supergravity on Riemann-flat manifolds that lead to classical Minkowski vacua with fully broken supersymmetry, we can see why the supertrace relations we found above have a general validity. 
For simplicity we will work in the language of $D=4$ non-compact dimensions, but our considerations will be valid for any number of compact and non-compact flat dimensions.

Assume that we can define up to seven U(1)$_A$ factors  ($A=1,\ldots,n$; $n \le 7$) under which the eight four-dimensional supercharges $Q_i$ ($i=1,\ldots,8$) transform non-trivially, and denote by $\vec{q}_i \equiv \{ q_i^A \}_{A=1,\ldots,n}  \equiv (q_i^1,\ldots,q_i^n)$ the vector of the charges of $Q_i$ with respect to U(1)$^n$. 
In four -dimensional language, the 2 gravitons of helicity $\pm2$, the 8 gravitinos of helicity $\pm 3/2$, the 28 gauge bosons of helicity $\pm1$, the 56 fermions of helicity $\pm 1/2$ (including the goldstinos) and the 70 scalars of helicity 0 (including the Goldstone bosons) are obtained by starting from the  graviton of helicity $+ 2$ and by acting repeatedly with all the available supercharges. This uniquely determines the charges of all the states, according to the following scheme:
\begin{eqnarray}
| +2 \rangle : & & \vec{0} \, , 
\nonumber \\
| +3/2 \, , \, i \rangle =  Q_i \, | +2 \rangle : & & \vec{q}_i \, , 
\nonumber \\
| +1 \, , \, [ij] \rangle =  Q_i \, Q_j | +2 \rangle : & & \vec{q}_i + \vec{q}_j \, , 
\nonumber \\
| +1/2 \, , \, [ijk] \rangle =  Q_i \, Q_j \, Q_k | +2 \rangle : & & \vec{q}_i + \vec{q}_j + \vec{q}_k \, , 
\nonumber \\
| 0 \, , \, [ijkl] \rangle =  Q_i \, Q_j \, Q_k \, Q_l | +2 \rangle : & & \vec{q}_i + \vec{q}_j + \vec{q}_k + \vec{q}_l \, , 
\label{U1charges}
\end{eqnarray}
and similarly for the CPT-conjugates. Acting with all 8 supercharges on the graviton of helicity $+2$ must give back the graviton of helicity $-2$, which leads to the constraint:
\begin{equation}
\sum_{i=1}^8 \vec{q}_i = \vec{0} \, . 
\label{CPT}
\end{equation}
Our second assumption is that the KK spectrum can be represented as follows:
\begin{eqnarray}
|2 \rangle : & & M^2 = \vec{n}^2 \, , 
\nonumber \\
|3/2 \, , \, i \rangle : & & M_i^2 = \left( \vec{n} + \vec{q}_i \right)^2 \, , 
\nonumber \\
|1 \, , \, [ij] \rangle : & & M_{ij}^2 = \left( \vec{n} + \vec{q}_i + \vec{q}_j \right)^2  \, , 
\nonumber \\
|1/2 \, , \, [ijk] \rangle : & & M_{ijk}^2 = \left( \vec{n} + \vec{q}_i + \vec{q}_j + \vec{q}_k \right)^2  \, , 
\nonumber \\
| 0 \, , \, [ijkl] \rangle : & & M_{ijkl}^2 = \left( \vec{n} + \vec{q}_i + \vec{q}_j + \vec{q}_k + \vec{q}_l \right)^2 \, , 
\label{U1spectrum}
\end{eqnarray}
where the scalar products of the charge vectors must be taken with a suitable field-dependent real diagonal metric.
Setting $\vec{M}_a = \{ \vec{n}, \vec{n}+\vec{q}_i, \vec{n}+\vec{q}_i+\vec{q}_j,   \vec{n}+\vec{q}_i+\vec{q}_j +\vec{q}_k ,   \vec{n}+\vec{q}_i+\vec{q}_j+\vec{q}_k +\vec{q}_l  \}$:
\begin{equation}
(\vec{M}_a)^2 =  \sum_{A=1}^n (M_a^A)^2 \, \mu_A^2 \, .
\label{U1metric}
\end{equation}
Implicitly, we are working in a gauge where the goldstinos, which provide the $\pm 1/2$ helicity components of the massive gravitinos, are assigned masses equal to the corresponding gravitino masses, and the Goldstone bosons, which provide the longitudinal component of the massive vectors or tensors, are assigned masses equal to the corresponding vector or tensor boson masses. This simplifies the computation of the supertraces, since we can now sum also over unphysical states (goldstinos and Goldstone bosons) and write:
\begin{equation}
{\rm Str} \, {\cal M}^{2p} =
{\rm Tr} \, [{\cal M}_{(0)}^2]^p - 2 \ 	{\rm Tr} \, [{\cal M}_{(1/2)}^2]^p + 2 \ {\rm Tr} \, [{\cal M}_{(1)}^2]^p - 2 \ {\rm Tr} \, [{\cal M}_{(3/2)}^2]^p + {\rm Tr} \, [{\cal M}_{(2)}^2]^p
 \, .
\label{Supertracerev}
\end{equation}
It is immediate to check that, under the two assumptions specified above,
\begin{equation}
{\rm Str} \, {\cal M}^2 = {\rm Str} \, {\cal M}^4 = {\rm Str} \, {\cal M}^6 =  0 \, , 
\label{m6zer}
\end{equation}
and, after setting $\epsilon_a = 1, -2,2,-2,2$ for the 70 spin-0, the 56 spin-1/2, the 28 spin-1, the 8 spin-3/2 and the spin-2:
$$
{\rm Str} \, {\cal M}^8 =  \sum_a \epsilon_a \left(  \vec{M}_a^{\; 2} \right)^k =  \sum_a \epsilon_a \left[  \sum_{A} \left( M_a^A \right)^2 \, \mu_A^2 \right]^k \, .
$$
If there is only a single U(1)  under which the supersymmetry generators are charged, as in the case of the mechanism of ref.~\cite{Cremmer:1979uq} on the circle:
\begin{equation}
{\rm Str} \, {\cal M}^8 =  40320 \, \left( \prod_{i=1}^8 q_i \right) \, \mu^8 \, .
\end{equation} 
If instead the charges $q_i^A$ do not depend on $A$, then, as in the case of $d=3$ HW:
\begin{equation}
{\rm Str} \, {\cal M}^8 =  40320 \, \left( \prod_{i=1}^8 q_i \right) \, \left( \sum_A \mu_A^2  \right)^4 \, .
\end{equation} 



\section{Flat manifolds and Scherk--Schwarz truncations} 
\label{sec:flat_manifolds_and_scherk_schwarz_reductions}

The computation of the 1-loop potential in the previous section is an improvement of the calculation in the reduced theory, including the contribution of the KK states.
All the examples we are considering are fully broken models coming from a maximally supersymmetric theory, where the supersymmetry breaking is spontaneous and hence one recovers full supersymmetry at the Lagrangian level on any consistent reduced theory.

Looking at properties of the different possible Riemann-flat manifolds,  we can distinguish two main classes of compactifications.
These two classes depend on the existence of submaximal orbits of the holonomy group and are nicely represented by the two main examples discussed so far, the ${\mathbb T}_3/{\mathbb Z}_3$  and the HW manifolds.
In fact, in the case of ${\mathbb T}^3/{\mathbb Z}_3$ (the same is true for any other ${\mathbb Z}_k$ freely acting orbifold), we can obtain a reduced theory from a Scherk--Schwarz reduction along the lines discussed in \cite{DallAgata:2005zlf,Grana:2013ila}, while for the HW manifold this cannot be achieved.
The first class is indeed such that the same flat manifold can also be realized as a quotient of a non-compact group-manifold whose generators are the same as those defining the lattice, while this is not true in general.
The global Scherk--Schwarz twist matrices lead to identifications on the globally defined vielbeins that preserve certain isometries or not.
In fact, as noticed in \cite{DallAgata:2005zlf,Grana:2013ila}, different identifications in the Scherk--Schwarz truncations result in the matching of the reduction with a truncation of the compactification on the Bieberbach flat manifold or just on its covering space, the $d$-dimensional torus.

Following \cite{Grana:2013ila}, globally defined vielbeins on ${\mathbb T}^3/{\mathbb Z}_k$ are given by 
\begin{equation}
		e^a = dy^i e_i{}^a  = \left(\begin{array}{c}
		\cos (qz)dy^1- \sin(qz) dy^2 \\
		\sin (qz) dy^1+ \cos(qz) dy^2 \\
		dy^3
		\end{array}\right)
\end{equation}
where $q = {2 \pi} k p$ or $q = {2 \pi} k p+ 2 \pi$, for $p \in {\mathbb Z}$ (the value of $q$ has been renormalized to match the shift by $1/k$ rather than unit).
In the first case we end up with the flat torus, in the second case with the Bieberbach manifold.
The difference lies in the fact that in both cases we want to constrain ourselves to singlets of the underlying group manifold, but in the first case we move along the lattice with unit steps and in the second we really build a freely acting orbifold, moving in the $y^3$ direction with a step of $1/k$.
The gauged supergravity obtained by the corresponding twisted torus reduction in the first case then is a consistent truncation of the spectrum of the torus, the second is a consistent truncation of the spectrum of the Bieberbach manifold.

Since we have been able to compute the full KK spectrum on this manifold, we can explicitly see the correspondence.
Before proceeding, we should note that the two reductions have a profound difference in the extent of their application.
While the Scherk--Schwarz reduction gives a full non-linear reduction scheme and the final result provides a consistent truncation of the original higher-dimensional theory, the KK spectrum is determined by the linear fluctuations about a higher-dimensional background.
For this reason, the two reductions can be matched only after identifying the full non-linear modes selected in Scherk--Schwarz and linearizing the reduction ansatz.
The other important point is that the two descriptions use different Lorentz gauges to patch the manifold. 
Using the globally defined vielbeins to perform the expansion and the non-linear Schwarz--Schwarz reduction, the group structure becomes the identity and this can be inherited on the spin structure by the expansion with globally defined constant spinors, which do not necessarily exist in the gauge associated to the KK expansion explained before.

The first observation is that if we reassemble the globally defined vielbeins as 
\begin{equation}\label{identif}
		e^1 + i e^2 = \exp\left(i q z \right)(dy^1 + i dy^2), \quad e^3 = dy^3,
\end{equation}
we match the 1-form harmonics in the short series (\ref{harm1}).
This allows us to identify and match the states of the two reduction schemes.
Let us start from the differential forms.
For the 1-form $A$, the Scherk--Schwarz ansatz selects the states
\begin{equation}
	A(x,y) = dy^i a_i(x) + e^a(y) a_a(x) = a_7(x) + e^a(y) a_a(x).
\end{equation}
Noting that 
\begin{equation}
	d e^1 = q\, e^2 \wedge e^3, \quad d e^2 = - q\, e^1 \wedge e^3, \quad de^3 = 0,
\end{equation}
the action of the higher-dimensional Laplacian on the second term in the expansion, generates $q^2$ masses for the fields $a_1(x)$ and $a_2(x)$.
It is quite straightforward to realize that these same massive states appear in the ordinary KK reduction, if we perform the comparison
\begin{equation}
	e^a(y) a_a(x) = dy^i e_{i}^a(y) a_a(x) = dy^{i} Y_{i}^I(y) a_I(x),
\end{equation}
for $I$ the vector harmonic with $m=n=0$.
This same match works for the states associated to $B_{i i}$, $C_{i\nu i}$ and $g_{i i}$.
All the massless states in the Scherk--Schwarz reduction, like the 1-form $a_7(x)$, are associated to the constant scalar harmonic, or to the closed, non-exact forms on the Bieberbach manifold.
Finally, for the bosonic states, we need to relate the states that come from the expansion of the metric with the symmetric harmonic.
Also in this case the match works by comparing the Scherk--Schwarz reduction for the expansion of the internal manifold fluctuations
\begin{equation}
	ds_3^2 = e^a(y) \otimes e^b(y) g_{ab}(x),
\end{equation}
with the KK one
\begin{equation}
	ds_3^2 = dy^i \otimes dy^j \, S_I(x) Y^I_{i j}(y),
\end{equation}
where $Y_{ij}^I$ are the symmetric tensor harmonics.
The transverse symmetric harmonics do not contain $Y_{33}$, nor $Y_{31}$ or $Y_{32}$ components.
However, one should note that the Scherk--Schwarz states $g_{31}$ and $g_{32}$ are goldstone modes.
This is clear once we linearize the expansion of the covariant derivatives on the metric scalars about the background
\begin{equation}
	D_i g_{i j} = \partial_i g_{i j} - g_{i k} \tau^{k}_{jk} g_i{}^k-g_{j k} \tau^{k}_{i k} g_i{}^k,
\end{equation}
where $\tau_{23}^1 = q$ and $\tau_{13}^2 = -q$, as follows from the derivatives of the background vielbeins.
Once we take linear fluctuations we get that 
\begin{eqnarray}
	D_i g_{31} &=& \partial_i g_{31} + q g_i{}^2, \\[2mm]
	D_i g_{32} &=& \partial_i g_{32} - q g_i{}^1.
\end{eqnarray}
The $g_{33}$ mode is the additional massless mode coming from the existence of non-trivial 1-cycles on the manifold.
Finally, the remaining combinations $g_{ww}$ and $g_{w \bar{w}}$, where $w = y^1+i\, y^2$ give the additional scalar states with masses $\pm 2$ and $0$ in the Table (\ref{tabellamasse}), once we pick the symmetric tensor harmonics with $m=n=0$ and appropriate $p$, according to what was chosen for the vielbeins.

A similar identification follows for the fermions, though we have to deal with some subtleties, which we will now describe.

Let us start with the spin 1/2 fields.
The fermion expansion in the Scherk--Schwarz reduction assumes we are expanding about the vielbein basis, rather than the coordinate one.
The internal Dirac action is then
\begin{equation}\label{Dirac}
		{\cal L}_{1/2} =\bar{\psi}\sigma^a e_a{}^i \left(\partial_{i} + \frac14 \omega_{i}{}^{bc} \sigma_b \sigma_c\right)\psi,
\end{equation}
where the connection follows from the vielbein definition above
\begin{equation}
		de^a + \omega^a{}_b e^b = 0
\end{equation}
and is
\begin{equation}
		\omega_3{}^{12} = q.
\end{equation}
This gives the following explicit form for the action
\begin{equation}
		{\cal L}_{1/2} =\bar{\psi}\left[\sigma^3 \partial_3 \psi + (\cos (qy^3) \sigma^1 + \sin(qy^3) \sigma^2)\partial_1 \psi + (\cos (qy^3) \sigma^2 - \sin(qy^3) \sigma^1)\partial_2 \psi + \frac{i}{2} q \psi\right],
\end{equation}
which, for a constant $\psi$, gives an effective mass of magnitude $q/2$.
However, before matching the state to the KK expansion it is interesting to rewrite  (\ref{Dirac}) in the suggestive form
\begin{equation}
		\begin{split}
				{\cal L}_{1/2} =&\bar{\psi}\sigma^3 \partial_3 \psi + e^{\left(-\frac{i}{2} q y^3 \sigma^3\right)} \sigma^1 e^{\left(\frac{i}{2} q y^3 \sigma^3\right)}\partial_1 \psi + e^{\left(-\frac{i}{2} q y^3 \sigma^3\right)} \sigma^2 e^{\left(\frac{i}{2} q y^3 \sigma^3\right)}\partial_2 \psi + \frac{i}{2} q \psi  \\
				=&\bar{\psi}e^{\left(-\frac{i}{2} q y^3 \sigma^3\right)}  \sigma^{i} \partial_{i}\left[ e^{\left(\frac{i}{2} q y^3 \sigma^3\right)} \psi\right].
		\end{split}
\end{equation}
This tells us that the equation of motion in the vielbein basis is related to the one in coordinate basis by using the spinor
\begin{equation}
		\Xi =  e^{\left(\frac{i}{2} q \,y^3\, \sigma^3\right)} \psi,
\end{equation}
for a constant $\psi$.
We therefore see that the requirement that a globally defined constant spinor exists in the Scherk--Schwarz reduction translates into the existence of globally defined spinors $\Xi$ in the KK reduction.
It is interesting that depending on the choice of $p$ such spinors are indeed associated to different spinor structures on the Bieberbach manifold and hence also to different KK spectra.
Sending $y^3 \to y^3+1$ gives
\begin{equation}
	\Xi \to e^{i \pi\, (k p + 1)} \Xi,
\end{equation}
and hence one gets periodic or antiperiodic conditions according to $k$ and $p$.
Still we can always match the spectrum with the KK one by using the corresponding fermionic structure as in \cite{Pfaffle}, which gives 
\begin{equation}\label{mperiodic}
	p = 2 \pi \left(k {\mathbb Z} + \frac{k+1}{2}\right)
\end{equation}
for $\delta=-1$, which gives periodic conditions for $k=3$ and antiperiodic for $k=2,4,6$ and
\begin{equation}\label{mtwisted}
	p = 2 \pi\left(k{\mathbb Z} + \frac12\right)
\end{equation}
for $\delta =1$, which gives twisted conditions for any allowed $k$.
The match is straightforward if we identify $p = 2{\mathbb Z}+1$ or $p = 2 {\mathbb Z}$ and we plug the result in $m = q/2$.

For the 3-dimensional case at hand the internal gravitino action becomes
\begin{equation}
	{\cal L}_{3/2} = \epsilon^{abc} e_b{}^{i} \bar{\psi}_a \left(\partial_{i} + \frac14\, \omega_{i}{}^{ef} \sigma_e \sigma_f\right)\psi_c + \epsilon^{abc} e_b{}^i \omega_{i c}{}^d  \bar{\psi}_a\psi_d.
\end{equation}
By taking appropriate complex combinations $\psi_w = \psi_1+ i \, \psi_2$, we can see that the equations of motion reduce to 
\begin{eqnarray}
	\partial_3 \psi_w + i\,q\,\left({\mathbb 1}+ \frac{\sigma_3}{2}\right)\psi_w = 0,
\end{eqnarray}
for the harmonics with $\psi_z =0$.
This equation can also be recast in the form
\begin{eqnarray}
	 \partial_3 \left(e^{i\, q\, \left({\mathbb 1}+ \frac{\sigma_3}{2}\right)y^3}\psi_w\right) = 0,
\end{eqnarray}
which suggests once more the identification of the Scherk--Schwarz twisted states with
\begin{equation}
	\Xi_w = e^{i\, q\, \left({\mathbb 1}+ \frac{\sigma_3}{2}\right)y^3}\psi_w,
\end{equation}
for $\psi_w$ constant, in the KK expansion.

According to our discussion of the spin 3/2 fields expansion, these are actually longitudinal harmonics and indeed they once more can be matched with the derivatives of the spin 1/2 harmonics.


%

\bigskip

\subsection*{Acknowledgments} 
\label{sub:acknowledgments}

This work was supported in part by the Italian MUR Departments of Excellence grant
2023-2027 ``Quantum Frontiers'' and by the MUR-PRIN contract 2022YZ5BA2 - ``Effective
quantum gravity''.
We thank A.~Kehagias, G.~Inverso, R. Lutowski, M.~Montero and T.~Van Riet for useful discussions.
G.~D. thanks also the organizers and participants of the workshop ``Conference on Geometry, Strings and the Swampland Program'', March 18 - 22, 2024, Ringberg Castle, Tegernsee, where preliminary results of this work were presented and discussed.


\begin{appendix}
	\section{Properties of the Epstein Zeta function} 
	\label{sec:properties_of_the_epstein_zeta_function}
	
	The Epstein zeta function \cite{epsteinzeta} associated with a $d$-dimensional lattice $\Lambda$  is defined by 
	\begin{equation}\label{epsdef}
		Z_{\Lambda}[s,\vec{a},\vec{b}^*] = \sum_{\vec z \in \Lambda} \frac{e^{-2 \pi i \vec{b}^*\cdot\vec{z}}} {\lVert\vec{z} + \vec{a}\rVert^s},
	\end{equation}
	where $\vec{z} = z_a \vec{e}_a \in \Lambda$ (i.e. $z_a \in {\mathbb Z}$), $\vec{a} = a_a \vec{e}_a$, $\vec{b}^* = b_a \vec{e}_a^{\, *}$, with $a_a, b_a \in {\mathbb R}$, for $\vec{e}_a$ and $\vec{e}_a^{\, *}$ basis for $\Lambda$ and $\Lambda^*$ respectively, and the sum skips $\vec{z} = -\vec{a}$.
	Some useful properties are listed below.
		\begin{enumerate}
		\item Reflection symmetry:	\begin{equation}
			Z_{\Lambda}[s,-\vec{a},\vec{b}^*] = Z_{\Lambda}[s,\vec{a},-\vec{b}^*],
	\end{equation}
		\item Another reflection symmetry:	\begin{equation}
			Z_{\Lambda}[s,-\vec{a},-\vec{b}^*] = Z_{\Lambda}[s,\vec{a},\vec{b}^*],
	\end{equation}
	\item If any $b_a$ is increased by an integer, the right hand side of (\ref{epsdef}) is unaffected.
	\item If $a_a$ is increased by unity, the right hand side of (\ref{epsdef}) is multiplied by $\exp(-2 \pi i \, b_a)$.
	\item It obeys the functional equation 
	\begin{equation}
e^{-2\pi i \vec{a}^{\, *} \cdot\vec{b}} \frac{\Gamma\left(\frac{d-x}{2}\right)}{\pi^{\frac{d-x}{2}}} Z_{\Lambda^*}[d-x,\vec{a}^{\, *},-\vec{b}] = 	\sqrt{\Delta}\,		\frac{\Gamma(x/2)}{\pi^{x/2}} Z_{\Lambda}[x,\vec{b},\vec{a}^{\, *}] ,
	\end{equation}
	with $\Delta$ the determinant of the Gram matrix defining the norm for the lattice $\Lambda$.
	\item It is vanishing for negative even integers:\begin{equation}
			Z_{\Lambda}[-2n,\vec{a},\vec{b}^*] = 0, \qquad \forall n\in {\mathbb N} \, .
	\end{equation}
	\item It generically vanishes at the origin 
	\begin{equation}
			Z_{\Lambda}[0,\vec{a},\vec{b}^*] = 0, 
	\end{equation} 
	unless $\vec{a} \in \Lambda$, in which case
	\begin{equation}
				Z_{\Lambda}[0,\vec{a},\vec{b}^*] = - \exp(-2 \pi i \vec{a}\cdot\vec{b}^*).
		\end{equation}
	\item If $a_a$ and $b_a$ are 0 or 1/2 and $4 a_a b_a$ is odd, then (\ref{epsdef}) is identically zero, for all $s$ and $\Lambda$.
	\end{enumerate}
	
When the lattice $\Lambda$ is 1-dimensional, $d=1$, the sum reduces to the Hurwitz zeta function
\begin{equation}
		Z_{\mathbb Z}[s,a,0] = 2 \zeta(s,a),
\end{equation}
satisfying the useful identities
\begin{equation}
		\pi^\frac{s-1}{2}\, \Gamma\left(\frac{1-s}{2}\right)\, \zeta(1-s) = \frac{\Gamma(s/2)}{\pi^{s/2}} \zeta(s),
\end{equation}
and
\begin{equation}
		\sum_{p=1}^n \zeta(s,p/n) = (n^s-1)\zeta(s).
\end{equation}

\end{appendix}




\bigskip


\begin{thebibliography}{100}
	
\bibitem{Scherk:1978ta} J.~Scherk and J.~H.~Schwarz, {\em Spontaneous Breaking of Supersymmetry Through Dimensional Reduction}, Phys. Lett. B \textbf{82} (1979), 60-64 doi:10.1016/0370-2693(79)90425-8
%
\bibitem{Scherk:1979zr} J.~Scherk and J.~H.~Schwarz, {\em How to Get Masses from Extra Dimensions}, Nucl. Phys. B \textbf{153} (1979), 61-88 doi:10.1016/0550-3213(79)90592-3
%
\bibitem{Cremmer:1979uq} E.~Cremmer, J.~Scherk and J.~H.~Schwarz, {\em Spontaneously Broken N=8 Supergravity}, Phys. Lett. B \textbf{84} (1979), 83-86 doi:10.1016/0370-2693(79)90654-3
%


\bibitem{Zwirner:2025ohv}
F.~Zwirner,
{\em No-scale supergravity},
[arXiv:2504.06190 [hep-th]].


\bibitem{DallAgata:2005zlf} G.~Dall'Agata and N.~Prezas, {\em Scherk--Schwarz reduction of M-theory on G2-manifolds with fluxes}, JHEP \textbf{10} (2005), 103 doi:10.1088/1126-6708/2005/10/103 [arXiv:hep-th/0509052 [hep-th]].


%
\bibitem{Grana:2013ila} M.~Gra{\~n}a, R.~Minasian, H.~Triendl and T.~Van Riet,
{\em Quantization problem in Scherk--Schwarz compactifications},'
Phys. Rev. D \textbf{88} (2013) no.8, 085018 doi:10.1103/PhysRevD.88.085018 [arXiv:1305.0785 [hep-th]].

\bibitem{Dixon:1986iz} L.~J.~Dixon and J.~A.~Harvey, {\em String Theories in Ten-Dimensions Without Space-Time Supersymmetry}, Nucl. Phys. B \textbf{274} (1986), 93-105 doi:10.1016/0550-3213(86)90619-X
%
\bibitem{Alvarez-Gaume:1986ghj} L.~Alvarez-Gaume, P.~H.~Ginsparg, G.~W.~Moore and C.~Vafa, {\em An $O(16) \times O(16)$ Heterotic String}, Phys. Lett. B \textbf{171} (1986), 155-162 doi:10.1016/0370-2693(86)91524-8
%
\bibitem{Rohm:1983aq} R.~Rohm, {\em Spontaneous Supersymmetry Breaking in Supersymmetric String Theories}, Nucl. Phys. B \textbf{237} (1984) 553 doi:10.1016/0550-3213(84)90007-5
%
\bibitem{Ferrara:1987es} S.~Ferrara, C.~Kounnas and M.~Porrati, {\em Superstring Solutions With Spontaneously Broken Four-dimensional Supersymmetry}, Nucl. Phys. B \textbf{304} (1988), 500-512 doi:10.1016/0550-3213(88)90639-6
%
\bibitem{Kounnas:1988ye} C.~Kounnas and M.~Porrati, {\em Spontaneous Supersymmetry Breaking in String Theory}, Nucl. Phys. B \textbf{310} (1988), 355-370 doi:10.1016/0550-3213(88)90153-8
%
\bibitem{Ferrara:1988jx} S.~Ferrara, C.~Kounnas, M.~Porrati and F.~Zwirner, {\em Superstrings with Spontaneously Broken Supersymmetry and their Effective Theories}, Nucl. Phys. B \textbf{318} (1989), 75-105 doi:10.1016/0550-3213(89)90048-5
%
\bibitem{DallAgata:2024ijh} G.~Dall'Agata and F.~Zwirner, {\em One-Loop Effective Potential in Scherk--Schwarz Compactifications of Pure $d=5$ Supergravities}, Fortsch. Phys. \textbf{72} (2024) no.7-8, 2400087 doi:10.1002/prop.202400087 [arXiv:2401.02480 [hep-th]].

\bibitem{DallAgata:2012tne} G.~Dall'Agata and F.~Zwirner, {\em Quantum corrections to broken N = 8 supergravity}, JHEP \textbf{09} (2012), 078
doi:10.1007/JHEP09(2012)078 [arXiv:1205.4711 [hep-th]].

\bibitem{Andriot:2018tmb}
D.~Andriot and D.~Tsimpis,
{\em Laplacian spectrum on a nilmanifold, truncations and effective theories},
JHEP \textbf{09} (2018), 096
doi:10.1007/JHEP09(2018)096
[arXiv:1806.05156 [hep-th]].

\bibitem{Gkountoumis:2023fym} G.~Gkountoumis, C.~Hull, K.~Stemerdink and S.~Vandoren, {\em Freely acting orbifolds of type IIB string theory on T$^{5}$}, JHEP \textbf{08} (2023), 089 doi:10.1007/JHEP08(2023)089 [arXiv:2302.09112 [hep-th]].

\bibitem{Acharya:2020hsc} B.~S.~Acharya, G.~Aldazabal, E.~Andr{\'e}s, A.~Font, K.~Narain and I.~G.~Zadeh, {\em Stringy Tachyonic Instabilities of Non-Supersymmetric Ricci Flat Backgrounds},'JHEP \textbf{04} (2021), 026 doi:10.1007/JHEP04(2021)026 [arXiv:2010.02933 [hep-th]].

\bibitem{epsteinzeta} J.~M.~Borwein, M.~L.~Glasser,  R.~C.~McPhedran, J.~G.~Wan and I. ~J.~Zucker, {\em Lattice sums then and now}, Encyclopaedia of Mathematics and its Applications series, CUP (2013). https://doi.org/10.1017/cbo9781139626804.

\bibitem{Bento} B.~V.~Bento and M.~Montero, {\em An M-theory dS maximum from Casimir energies on Riemann-flat manifolds}, [arXiv:2507.02037 [hep-th]].

\bibitem{Charlap1986} L.S.~Charlap, {\em Bieberbach Groups and Flat Manifolds}, Universitext, Springer-Verlag New York Inc. 1986.
doi:10.1007/978-1-4613-8687-2

\bibitem{Kirby} R. C. Kirby, {\em The topology of 4-manifolds}, Lecture Notes in Mathematics 1374, Springer-Verlag (1989).

\bibitem{Friedrich} Th. Friedrich, {\em Dirac Operators in Riemannian Geometry}, Graduate
Studies in Mathematics No. 25, AMS 2000.

\bibitem{Cid}  C.~Cid, T.~Schulz, {\em Computation of Five- and Six-Dimensional Bieberbach Groups},  Experiment. Math. 10(1): 109-115 (2001). 

\bibitem{Lutowski} R. Lutowski, B. Putrycz, {\em Spin structures on flat manifolds}, Journal of Algebra 436 (2015) 277-291 [arXiv:1411.7799]
doi:10.1016/j.jalgebra.2015.03.037.
	
\bibitem{Handtsche1935} W. Hantzsche, H. Wendt, {\em Dreidimensionale euklidische Raumformen}, Mathematische Annalen 110 (1935), 593–611.


\bibitem{Riazuelo:2003ud}
A.~Riazuelo, J.~Weeks, J.~P.~Uzan, R.~Lehoucq and J.~P.~Luminet,
{\em Cosmic microwave background anisotropies in multi-connected flat spaces},
Phys. Rev. D \textbf{69} (2004), 103518
doi:10.1103/PhysRevD.69.103518
[arXiv:astro-ph/0311314 [astro-ph]].


\bibitem{Peng:2019xoj} Z.~P.~Peng, L.~Lindblom and F.~Zhang, {\em Scalar, Vector and Tensor Harmonics on the Flat Compact Orientable Three-Manifolds}, JCAP \textbf{12} (2019), 042 doi:10.1088/1475-7516/2019/12/042 [arXiv:1909.06721 [gr-qc]].

\bibitem{Kehagias:2004gy} A.~Kehagias and K.~Tamvakis, {\em Box compactification and supersymmetry breaking},'Phys. Lett. B \textbf{603} (2004), 249-256  doi:10.1016/j.physletb.2004.10.019 [arXiv:hep-th/0403029 [hep-th]].

\bibitem{Pfaffle} F.~Pf\"affle, {\em The Dirac spectrum of Bieberbach manifolds}, J. Geom. Phys. 35 (367–385), 2000.

\bibitem{Miatello2001} R. J.~Miatello and J. P.~Rossetti, {\em Flat manifolds isospectral on p-forms}, Jour. Geom. Anal. 11 (647–665), 2001. [arXiv:math/0303276]
doi:10.1007/BF02930761.
	
\bibitem{Miatello2003} R.J.~Miatello and R.A.~Podest{\'a}, {\em The spectrum of twisted Dirac operators on compact flat manifolds}, Transactions of the American Mathematical Society 358, 4569-4603 (2003) [arXiv:math/0312004]
doi:10.1090/S0002-9947-06-03873-6

\bibitem{Ponton:2001hq} E.~Ponton and E.~Poppitz, {\em Casimir energy and radius stabilization in five-dimensional orbifolds and six-dimensional orbifolds}, JHEP \textbf{06} (2001), 019 doi:10.1088/1126-6708/2001/06/019 [arXiv:hep-ph/0105021 [hep-ph]]. 

\bibitem{Dienes:1995pm}
K.~R.~Dienes, M.~Moshe and R.~C.~Myers,
{\em String theory, misaligned supersymmetry, and the supertrace constraints},
Phys. Rev. Lett. \textbf{74} (1995), 4767-4770
doi:10.1103/PhysRevLett.74.4767
[arXiv:hep-th/9503055 [hep-th]].

\bibitem{Abel:2015oxa}
S.~Abel, K.~R.~Dienes and E.~Mavroudi,
{\em Towards a nonsupersymmetric string phenomenology},
Phys. Rev. D \textbf{91} (2015) no.12, 126014
doi:10.1103/PhysRevD.91.126014
[arXiv:1502.03087 [hep-th]].

\bibitem{Abel:2024twz}
S.~Abel, K.~R.~Dienes and L.~A.~Nutricati,
{\em New nonrenormalization theorem from UV/IR mixing},
Phys. Rev. D \textbf{110} (2024) no.12, 126021
doi:10.1103/PhysRevD.110.126021
[arXiv:2407.11160 [hep-th]].

\bibitem{Lutclass} Classification of Bieberbach manifolds up to dimension 6: https://mat.ug.edu.pl/~rlutowsk/bieblib/.

\bibitem{Duff:1986hr} M.~J.~Duff, B.~E.~W.~Nilsson and C.~N.~Pope, {\em Kaluza--Klein Supergravity}, Phys. Rept. \textbf{130} (1986), 1-142.
doi:10.1016/0370-1573(86)90163-8.

\bibitem{DAuria:1983rmi} R.~D'Auria and P.~Fre, {\em On the Fermion Mass Spectrum of {Kaluza--Klein} Supergravity},'Annals Phys. \textbf{157} (1984), 1.

\bibitem{Fabinger:2000jd} M.~Fabinger and P.~Horava, ``Casimir effect between world branes in heterotic M theory,'' Nucl. Phys. B \textbf{580} (2000), 243-263 doi:10.1016/S0550-3213(00)00255-8 [arXiv:hep-th/0002073 [hep-th]].

\bibitem{Bagger:2001ep} J.~Bagger, F.~Feruglio and F.~Zwirner, ``Brane induced supersymmetry breaking,'' JHEP \textbf{02} (2002), 010 doi:10.1088/1126-6708/2002/02/010 [arXiv:hep-th/0108010 [hep-th]].
%
\bibitem{Epsteinzetanum} A.~A.~Buchheit, J.~K.~Busse, R.~Gutendorf, {\em Computation and properties of the Epstein zeta function with high-performance implementation in EpsteinLib}, arXiv:2412.16317 [math.NA] doi:10.48550/arXiv.2412.16317.

\end{thebibliography}
\end{document}